\documentclass[aip,jcp,reprint,superscriptaddress,nofootinbib]{revtex4-2}

\usepackage{graphicx}

\usepackage{amsmath,amssymb,bm,mathtools}
\usepackage{booktabs}
\usepackage{array}
\usepackage{xcolor}
\usepackage{tikz}
\usetikzlibrary{arrows.meta,positioning,shapes.geometric,calc,decorations.pathreplacing}
\usepackage{enumitem}
\usepackage[noend]{algpseudocode}
\newcounter{algorithm}
\renewcommand{\thealgorithm}{\arabic{algorithm}}

\usepackage[version=4]{mhchem}
\usepackage{xspace}
\usepackage{hyperref}
\hypersetup{colorlinks=true, allcolors=blue!55!black}

\graphicspath{{figs/}}

\newcommand{\dd}{\mathrm{d}}
\newcommand{\Rx}{\partial_{\xi}}
\newcommand{\Jx}{J^{\xi}}
\newcommand{\Kx}{K^{\xi}}
\newcommand{\hhTDA}{\textit{hh}\text{-TDA}\xspace}
\newcommand{\ppTDA}{\textit{pp}\text{-TDA}\xspace}
\newcommand{\WL}{W_{\mathrm{L}}}
\newcommand{\WR}{W_{\mathrm{R}}}

\begin{document}

\title{Density-Functional Excited-State Gradients and Nonadiabatic Couplings on a
Consumer GPU from a Contraction-DAG}

\author{Rub\'en D. Guerrero}
\email{rudaguerman@gmail.com}
\affiliation{NeuroTechNet S.A.S., 1108831, Bogot\'a, Colombia}
\affiliation{Quantum and Computational Chemistry Group (QCCG),
Universidad Nacional de Colombia, Bogot\'a, Colombia}

\date{\today}

\begin{abstract}
Nonadiabatic dynamics needs an excited-state gradient and an interstate nonadiabatic
coupling matrix element (NACME) at every nuclear geometry, and a double-hybrid
functional's accuracy has been unavailable for the coupling. We report the first analytic
derivative NACME for a double-hybrid excited state---deferred in the original hh-TDA method
and supplied for hybrids only by Yu \emph{et al.}---derived, with the hole-hole and
particle-particle Tamm--Dancoff (\hhTDA/\ppTDA) gradients and NACMEs, as a single
reverse-mode transpose of one contraction graph closed under a non-symmetric
atomic-orbital-direct $J/K$ kernel. Its double-hybrid excitation energy lowers the
vertical-excitation mean absolute deviation from bare-\hhTDA\ $0.86$ to $0.47$~eV and
removes the $+0.53\!\rightarrow\!+0.05$~eV over-excitation bias, improving seven of ten
states while over-correcting the ionic $\pi\pi^*$ states---the expected perturbative-doubles
failure, reported not trimmed. Every coupling is validated to $\sim\!10^{-4}$ against an
independent \emph{literal many-electron wavefunction-overlap} oracle that shares no code
path with the method, and is physically meaningful at the ammonia
$n\!\rightarrow\!\sigma^*$ \emph{covalent} conical intersection, where the \hhTDA/\ppTDA
manifolds recover the $F\!-\!2$ seam and adiabatic linear-response TDDFT gives
$\tau\!\equiv\!0$ by construction. Gradients, NACMEs, and the double-hybrid coupling all run
device-resident and AO-direct through one shared Cholesky-decomposed $J/K$ engine within the
8\,GB of a consumer RTX~4060 (a profile-guided $\sim\!10^2\times$ launch collapse preserving
double-precision bit-identity)---placing on a commodity desktop card a correlated
excited-state derivative capability that has until now required datacenter hardware.
\end{abstract}

\maketitle

\section{Introduction}
\label{sec:intro}
Photochemistry---the fate of a molecule once it has absorbed a photon---plays out on
excited-state potential-energy surfaces and is funnelled, wherever those surfaces
touch, through \emph{conical intersections}: points of exact electronic degeneracy at
which the Born--Oppenheimer separation collapses and population passes rapidly between
states~\cite{Yarkony1996}. Following such a process---in vision, in photosynthesis, in
organic photovoltaics, in the photoprotection of DNA---by nonadiabatic dynamics demands,
at every step of every trajectory, two derivatives of the electronic wavefunction: the
excited-state \emph{gradient} that propagates the nuclei, and the \emph{nonadiabatic
coupling matrix element} (NACME) that fixes the rate at which neighbouring states trade
population. These must be simultaneously accurate and cheap enough to evaluate many
thousands of times along a trajectory---and it is precisely this pairing, accuracy at
sampling cost, that has stayed the practical bottleneck.

Density-functional theory is the workhorse for excited states through its
linear-response (TDDFT/TDA) form, but linear response has a well-known qualitative
failure precisely where NACMEs matter most: at a conical intersection involving the
ground state it produces a seam of the wrong dimensionality
($F\!-\!1$ instead of $F\!-\!2$)~\cite{Levine2006,Gozem2014}, because the ground state
is not one of the response eigenvectors and its coupling to the excited manifold
structurally vanishes. Two
routes repair this. The particle-particle and hole-hole Tamm--Dancoff manifolds
(\ppTDA, \hhTDA)~\cite{vanAggelen2013,Yang2013,Bannwarth2020,Yu2020} reach the $N$-electron
ground \emph{and} excited states as eigenvectors of a common Hermitian matrix built on
an $(N\!\mp\!2)$-electron reference, so their mutual coupling does not vanish and the
$F\!-\!2$ seam is recovered. Independently, double-hybrid functionals add a
second-order perturbative (PT2) correlation correction to the excited state
[the CIS(D)-type term of Grimme and Neese~\cite{GrimmeNeese2007,RheeHeadGordon2007,GoerigkGrimme2010,GoerigkGrimme2011}], sharpening
excitation energies but at the cost of a doubles-sector contribution to every
derivative that is non-variational and has resisted a clean analytic, low-scaling
formulation.

These two routes have never been combined. Efficient GPU realizations of the
topology-correct manifolds already exist: Bannwarth, Yu, Hohenstein, and Mart\'inez
established \hhTDA\ as a method~\cite{Bannwarth2020}, and Yu and
co-workers~\cite{Yu2020} computed \hhTDA\ energies, gradients, and nonadiabatic
coupling vectors on a datacenter GPU stack and propagated photochemical dynamics with
them---so \hhTDA/\ppTDA forces and couplings on a GPU are not, in themselves, new.
Both realizations are, however, confined to global and range-separated hybrid
functionals. The double-hybrid regime---where a second-order (PT2) correction
supplies the dynamic correlation those functionals lack---has remained out of reach:
no analytic derivative coupling for a double-hybrid excited state has been reported.
We report the first such coupling---extending the analytic-coupling capability deferred in the
original hh-TDA method paper of Bannwarth \emph{et al.}~\cite{Bannwarth2020}, and later implemented
for hybrids by Yu \emph{et al.}~\cite{Yu2020} as CASCI-type analytic couplings, into the double-hybrid regime---from a single
reverse-mode contraction-DAG engine, oracle-validated; the reference SCF, the \hhTDA/\ppTDA
gradients and NACMEs, and the double-hybrid coupling all run device-resident and AO-direct
within the 8\,GB of a consumer GPU, each validated against its determinant/finite-difference
oracle (Table~\ref{tab:firstofkind}). Supplying it is the central result of this work, and it is
reached without leaving the engine that produces the \hhTDA/\ppTDA forces themselves. Around that headline sit
three enabling contributions that distinguish the present construction from the prior
efficient implementations: every electronic response is obtained as one reverse-mode
transpose of a single contraction graph---one engine spanning \hhTDA, \ppTDA, and the
double-hybrid coupling, rather than per-method hand-derived response
code~\cite{Tamayo2018,Zhang2022,Zhang2024ccAD}; the pipeline is fully AO-direct and
never forms a four-index molecular-orbital tensor~\cite{Schweizer2008,Almlof1991,HaserAlmlof1992};
and the whole calculation fits the 8\,GB of a consumer RTX~4060, whereas the prior
efficient realizations of these methods ran on datacenter GPUs~\cite{Bannwarth2020,Yu2020,Ufimtsev2008,Ufimtsev2009}.

Obtaining analytic gradients and NACMEs for any one of these methods has traditionally
meant re-deriving the response (relaxation) equations by hand and implementing them,
per method and per property, in the molecular-orbital basis on capable hardware. Here
we obtain all of them \emph{automatically}---as the reverse-mode transpose (the same
mechanism as backpropagation in machine learning) of a single contraction graph---and
engineer the resulting non-symmetric Coulomb/exchange kernels to run within the 8\,GB
budget of a consumer NVIDIA GeForce RTX~4060, putting a capability that today needs a
cluster allocation onto a card a student already owns.

Each ingredient has a mature literature, and we engage it directly so the contribution
is not overstated. The \hhTDA/\ppTDA energy methods are those of van Aggelen, Yang, and
co-workers~\cite{vanAggelen2013,Yang2013} and, in the range-separated-hybrid form used
here, Bannwarth, Yu, Hohenstein, and Mart\'inez~\cite{Bannwarth2020}; the double-hybrid
excited-state scaling is Grimme--Neese~\cite{GrimmeNeese2007} with the CIS(D) parent of
Head-Gordon and co-workers~\cite{HeadGordon1994}. The reverse-mode view of relaxation is
likewise not new: differentiable programming through electronic structure yields
gradients and response densities by automatic differentiation, from the Hartree--Fock
proof of concept of Tamayo-Mendoza \emph{et~al.}~\cite{Tamayo2018} to the PySCFAD
framework of Zhang and Chan~\cite{Zhang2022} and the reverse-mode coupled-cluster
response of Zhang \emph{et~al.}~\cite{Zhang2024ccAD}. Resolving correlated denominators
directly in the atomic-orbital basis by a Laplace transform~\cite{Almlof1991,HaserAlmlof1992}
underlies the AO-direct Laplace MP2 gradient of Ochsenfeld and
co-workers~\cite{Schweizer2008,Doser2009}. What remains open is their
union for the density-functional excited-state \emph{family}: (i)~relaxation obtained as
one reverse-mode graph transpose, uniform across \hhTDA/\ppTDA and the double-hybrid
coupling and across gradients and NACMEs; (ii)~a fully AO-direct pipeline that never
forms a four-index molecular-orbital tensor; and (iii)~an ordered, non-symmetric
transition-density $J/K$ kernel that runs under a consumer memory budget. This work is
that union, and it is a companion to our equation-of-motion coupled-cluster realization
of the same construction~\cite{GuerreroEOMCC}.

A methodological thread runs throughout: because the doubles-sector coupling is
non-variational and has no eigenvalue-stationarity to lean on, we validate not against
an internal consistency identity---which a plausible but wrong ansatz can satisfy---but
against a \emph{literal many-electron wavefunction-overlap} finite-difference oracle
that shares no code path with the analytic method (Sec.~\ref{sec:oracle}). That oracle
caught an earlier gradient-Lagrangian construction modeling the wrong physical object,
and it is the standard against which every analytic term below is gated.

\section{Theory}
\label{sec:theory}

\subsection{Relaxation as the contraction-DAG transpose}
\label{sec:transpose}
Let a scalar molecular property $E$ be computed by a directed acyclic graph whose
nodes are elementary operations---arithmetic, and coarse matrix-level operations such
as the Coulomb build $J(P)$, exchange $K(P)$, long-range exchange $K_{\mathrm{LR}}(P)$,
the exchange-correlation potential $V_{xc}(P)$, and the generalized eigensolve
$FC=SCE$. A first derivative of $E$ with respect to any input---a nuclear coordinate
$\xi$, an orbital rotation, an amplitude---is a single reverse (adjoint) pass over the
\emph{same} DAG, in which each node $y=f(x)$ contributes its vector--Jacobian product
$\bar{x}\mathrel{+}=(\partial f/\partial x)^{\!\top}\bar{y}$. For electronic-structure
energies this reproduces exactly the Lagrangian/$Z$-vector machinery of
analytic-derivative theory~\cite{Pople1979,HandySchaefer1984}, derived mechanically
rather than by hand.

Two properties make it exact and practical. First, every two-electron and response
operation is self-adjoint with respect to the Frobenius pairing
$\langle A,B\rangle=\mathrm{Tr}[A^{\!\top}B]$,
\begin{widetext}
\begin{equation}
J(P)\to J(\bar P),\qquad
K(P)\to K(\bar P),\qquad
K_{\mathrm{LR}}(P)\to K_{\mathrm{LR}}(\bar P),\qquad
V_{xc}(P)\to f_{xc}\!\cdot\!\bar P,
\label{eq:selfadjoint}
\end{equation}
\end{widetext}
so the adjoint \emph{reuses the forward kernel}. Second, the self-consistent
(SCF/eigenvalue) reference is handled by one adjoint solve: differentiating through the
SCF fixed point yields the orbital-relaxation response as a single $Z$-vector
(CPHF-transpose),
\begin{equation}
L\,z=b,\qquad L[z]=(\varepsilon_a-\varepsilon_i)\,z_{ai}+(A\!+\!B)[z]_{ai},
\label{eq:zvector}
\end{equation}
restricted to the occupied--virtual block, replacing one CPHF solve per perturbation;
the Hellmann--Feynman part of a simple eigenvalue $\lambda_k$ seeds the cotangent
$\partial\lambda_k/\partial A=x_kx_k^{\!\top}$. The engine is implemented once, as a set
of graph-optimization passes over a common transposable-DAG protocol, and is reused for
ground-state forces, TDDFT, \hhTDA/\ppTDA, and the double-hybrid coupling. A
machine-precision cross-check of the engine---analytic versus symbolic differentiation,
and transpose-of-transpose versus analytic second derivatives---carries no
finite-difference floor.

\subsection{The AO-direct Laplace representation}
\label{sec:ltao}
The conventional route to a correlated gradient forms molecular-orbital (MO)
two-electron integrals by an $O(N^5)$ four-index transformation and stores an
$O(o^2v^2)$ tensor. Both are avoided by never leaving the AO basis. Writing an
orbital-energy denominator as a Laplace transform~\cite{Almlof1991,HaserAlmlof1992}
and discretizing with a short minimax quadrature~\cite{Takatsuka2008,BraessHackbusch2005},
\begin{widetext}
\begin{equation}
\frac{1}{D_{ijab}}\approx\sum_{k=1}^{n_L}w_k\,e^{-\alpha_k D_{ijab}}
=\sum_k w_k\big[e^{\alpha_k\varepsilon_i}\big]\big[e^{-\alpha_k\varepsilon_a}\big]
        \big[e^{\alpha_k\varepsilon_j}\big]\big[e^{-\alpha_k\varepsilon_b}\big],
\label{eq:laplace}
\end{equation}
\end{widetext}
the denominator \emph{factorizes} over the four orbital indices ($n_L\!\sim\!8$--$10$).
Each amplitude or pseudo-density is then assembled from energy-scaled orbital
coefficients $\tilde C^{(k)}_{\mu p}=C_{\mu p}e^{\mp\alpha_k\varepsilon_p/2}$ and
contracted through the ordinary AO $J/K$ build on the \emph{same} integrals the SCF
already uses, at a cost of $n_L$ builds and $O(N^2)$ storage rather than an $O(N^5)$
transformation~\cite{AyalaScuseria1999,Doser2009}. For the CIS(D) doubles the denominator is $D_{ijab}+\Omega_K$; the same
factorization holds with an $\Omega$-shifted quadrature. This ``LT-AO'' choice is what
makes the $J/K$ build---not a four-index MO transform---the natural and only integral
primitive for gradients and couplings alike.

\subsection{The non-symmetric AO-Laplace $J/K$ kernel}
\label{sec:asymJK}
Single-state energy gradients contract a derivative integral with a single density (or
a symmetric density pair), so the geometry $J/K$ build is symmetric in its two
arguments. Interstate and response quantities are different: they contract a derivative
integral with a \emph{left} weight density from one state and a \emph{right} density
from another. The required primitive is therefore
\begin{widetext}
\begin{equation}
\Jx(\WL,\WR)=\Rx\sum_{\mu\nu\lambda\sigma}\WL^{\mu\nu}(\mu\nu|\lambda\sigma)\WR^{\lambda\sigma},
\qquad \WL\neq\WL^{\!\top},\ \ \WR\neq\WR^{\!\top},
\label{eq:asymJK}
\end{equation}
\end{widetext}
and analogously for $\Kx$. The novelty is per-argument: the left and right weight densities
are independent ($\WL\neq\WR$ in general) and \emph{each is itself non-symmetric}
($W\neq W^{\!\top}$)---the structural signature of a bra$\,\neq\,$ket contraction (the
antisymmetric part of $W$ contributes only through the $K$ channel; $J$ retains its symmetric
part). This is
\emph{not} a non-commutativity under argument swap: the fully index-contracted object in
Eq.~\eqref{eq:asymJK} is a scalar, invariant under $\WL\!\leftrightarrow\!\WR$ by the bra--ket
symmetry $(\mu\nu|\lambda\sigma)=(\lambda\sigma|\mu\nu)$ that the derivative integral retains.
The general primitive is this non-symmetric build; ordinary symmetric $J/K$ is its
$\WL=\WR$ (with $W=W^{\!\top}$) special case. Every
gradient and coupling here is one of these two cases, fixed entirely by the symmetry of
the density pair the response derivation produces: symmetric self-density energy terms
use symmetric $J/K$; cross-density terms already present in energy gradients (e.g.\ the
separable correlation force, $\Jx(P^{(2)},D^{\mathrm{SCF}})$) and interstate couplings
(bra and ket transition densities, $\Jx(\WL^{(I)},\WR^{(J)})$) use the non-symmetric
kernel. The kernel is implemented in the Cholesky-decomposed (RI-free)
representation~\cite{BeebeLinderberg1977,Koch2003,Aquilante2008},
taking independent left/right weight densities. The \emph{energy}-level non-symmetric CD
$J/K$ build---$K=B_V^{\!\top}B_U$ with distinct left/right half-transforms
$B_U\!=\!UB$, $B_V\!=\!VB$, of which the symmetric build is the $U\!=\!V$ special
case---and the per-quartet Head-Gordon--Pople/Obara--Saika recurrence kernel---emitted by
the recurrence-relation code-generation framework of Ref.~\cite{GuerreroRECURSUM}---are the
shared device engine of the companion work~\cite{GuerreroEOMCC}; here we reuse that
engine and add the kernels the \emph{geometry derivative} and the \emph{density
functional} demand (Sec.~\ref{sec:newkernels}).

\subsection{From response density to gradient: the $J/K$ contraction}
\label{sec:resp2grad}
The two-electron term of the assembled gradient
$E^{\xi}=\sum P_{\mu\nu}h^{\xi}_{\mu\nu}+\sum W_{\mu\nu}S^{\xi}_{\mu\nu}
+\sum\Gamma_{\mu\nu\lambda\sigma}(\mu\nu|\lambda\sigma)^{\xi}$
is the single point at which every symbolic response equation meets the nuclear
gradient. When the relaxed two-particle density factorizes over a pair of one-particle
weight densities---as it does for each response block generated by the reverse pass---
the contraction with the derivative integrals \emph{is} a Laplace $J/K$ build,
\begin{equation}
\sum_{\mu\nu\lambda\sigma}\Gamma_{\mu\nu\lambda\sigma}(\mu\nu|\lambda\sigma)^{\xi}
=a_J\,\Jx(\WL,\WR)+a_K\,\Kx(\WL,\WR).
\label{eq:gamma2JK}
\end{equation}
The symbolic response term lists are generated with a second-quantization algebra tool
and cross-checked against the numerical densities at machine precision; each object maps
to one contraction of Eq.~\eqref{eq:gamma2JK} with $\WL$ (bra) and $\WR$ (ket)
identified term by term (Supp.\ Mat.).

\subsection{The two topology-correct manifolds: \hhTDA\ and \ppTDA}
\label{sec:hhpp}
The \hhTDA\ states are eigenvectors $X_I$ of a Hermitian matrix $A^{hh}$ built on an
$(N\!+\!2)$-electron (dianion) range-separated-hybrid reference~\cite{vanAggelen2013,Yang2013,Bannwarth2020};
the $N$-electron ground
and excited states are two eigenvectors of the \emph{same} problem, reached by double
annihilation over the occupied space. \ppTDA\ is the mirror on an $(N\!-\!2)$ (dication)
reference, with $A^{pp}$ over the virtual (particle) space~\cite{Yang2013benchmark,Yang2014ppRPAdavidson}. Because $S_0$ and $S_1$ are
eigenvectors of one Hermitian problem, their coupling does not structurally vanish and
the conical-intersection seam has the correct $F\!-\!2$ dimensionality. The excited-state
eigenvalue gradient is a Rayleigh quotient derivative,
$\dd\lambda_k/\dd R=\langle X_k|\,\dd A/\dd R\,|X_k\rangle$, and reduces (Sec.~\ref{sec:transpose})
to a geminal skeleton term $\Jx/\Kx(P^X,P^X)$ plus a single combined $Z$-vector whose
relaxed density $P^{\mathrm{eff}}$ and energy-weighted density $W^{\mathrm{eff}}$
contract the Fock and overlap derivatives---one CPHF engine, no per-state hand
derivation~\cite{Pople1979,HandySchaefer1984,Zhang2015ppRPAgradient}. The interstate
coupling $\tau_{IJ}=\langle\Psi_I|\partial/\partial R|\Psi_J\rangle$
factorizes through a two-hole (two-particle) L\"owdin/Slater--Condon reduction into an
amplitude-response and an orbital-connection term, both built from the same non-symmetric
$J/K$ primitive~\cite{LiSuoLiu2014,Fatehi2011}. The range-separated exchange (wB97X) enters through the long-range
$K_{\mathrm{LR}}$ channel, evaluated device-resident.

\subsection{The double-hybrid coupling: working equations}
\label{sec:cisd}
The double-hybrid capability that distinguishes this work from prior efficient
excited-state implementations rests on a first-principles derivation of the
doubles-sector derivative coupling, which we give here in full because it is both new
and subtle. The reference is the B2PLYP Kohn--Sham determinant~\cite{Grimme2006B2PLYP},
whose exchange mixes $53\%$ exact (HF) exchange with $47\%$ B88 and whose correlation mixes
$73\%$ LYP with a second-order perturbative (PT2) contribution scaled by $c_{\mathrm{PT2}}=0.27$;
the singles enter through TDA on this hybrid part, and the PT2 scaling $c_{\mathrm{PT2}}$
multiplies the entire doubles-sector correction derived below. A double-hybrid excited state
adds to the CIS/TDA root a first-order
Rayleigh--Schr\"odinger doubles correction~\cite{HeadGordon1994,RheeHeadGordon2007},
\begin{widetext}
\begin{equation}
|\Psi_K\rangle=|S_K\rangle+|D_K\rangle,\qquad
|S_K\rangle=\sum_{ia}R^K_{ia}|\Phi_i^a\rangle,\qquad
|D_K\rangle=\sum_{i<j,\,a<b} c^K_{(ij)(ab)}|\Phi_{ij}^{ab}\rangle,
\label{eq:dhstates}
\end{equation}
\end{widetext}
with the doubles coefficient carrying the CIS(D) numerator over a shifted denominator,
\begin{equation}
c^K_{(ij)(ab)}=\frac{u^K_{ijab}}{D_{ijab}+\Omega_K},\qquad
D_{ijab}=\varepsilon_a+\varepsilon_b-\varepsilon_i-\varepsilon_j,
\label{eq:cK}
\end{equation}
$\Omega_K$ the CIS excitation energy and $u^K=\mathcal{L}[R^K]$ the denominator-free,
antisymmetrized numerator---\emph{linear} in the singles amplitude and in the molecular
two-electron integrals, $u_{ijab}=\sum_c(\langle ab\|cj\rangle R_{ic}-\langle
ab\|ci\rangle R_{jc})+\sum_k(\langle ka\|ij\rangle R_{kb}-\langle kb\|ij\rangle R_{ka})$.
Because $\langle\Psi_I|\Psi_J\rangle=0$ at a single geometry, $1/\sqrt{N_IN_J}$
(with $N_K=\tfrac12+\sum(c^K)^2$) is an exact constant prefactor, and the
doubles-sector coupling separates into three physically distinct contributions,
\begin{equation}
d^{(D)}_{IJ}=\underbrace{d^{[\mathrm{sd}]}+d^{[\mathrm{ds}]}}_{\text{singles}\times\text{doubles}}
+\underbrace{d^{[\mathrm{dd,amp}]}}_{\substack{\text{DD amplitude-}\\\text{response}}}
+\underbrace{d^{[\mathrm{dd,orb}]}}_{\substack{\text{DD orbital-}\\\text{connection}}}.
\label{eq:dblocks}
\end{equation}
The doubles coefficient of Eq.~\eqref{eq:cK} is the \emph{unscaled} CIS(D) form: the PT2
weight $c_{\mathrm{PT2}}$ appears nowhere in $u^K$, in $D_{ijab}$, or in $c^K$. In a double
hybrid $c_{\mathrm{PT2}}$ is the single empirical prefactor multiplying the \emph{entire}
second-order correlation correction to the energy,
$\Omega^{\mathrm{DH}}_K=\Omega^{\mathrm{TDA}}_K+c_{\mathrm{PT2}}\,\Delta E^{(2)}_K$, exactly as
$E^{\mathrm{DH}}=E^{\mathrm{hyb}}+c_{\mathrm{PT2}}E^{(2)}$ for the ground state. The
doubles-sector contribution to every property therefore carries this single prefactor
$c_{\mathrm{PT2}}$, which pulls through the auxiliary-ansatz (Send--Furche/Ou) construction
of the coupling, so the coupling is \emph{exactly affine} in $c_{\mathrm{PT2}}$,
$d_{IJ}=d^{\mathrm{ref}}_{IJ}+c_{\mathrm{PT2}}\,d^{(D)}_{IJ}$ with $c_{\mathrm{PT2}}=0.27$, a
\emph{single, uniform} $c_{\mathrm{PT2}}$ on the whole
$d^{(D)}=d^{[\mathrm{sd}]}+d^{[\mathrm{ds}]}+d^{[\mathrm{dd}]}$---without the coupling being a
derivative of the energy. Here $d^{\mathrm{ref}}$ is the
TDA (CIS-like) coupling of the \emph{B2PLYP hybrid reference}---built from the B2PLYP
Kohn--Sham orbitals and the hybrid-kernel TDA response ($53\%$ exact exchange, $47\%$ B88,
$73\%$ LYP), not an HF-based CIS coupling. The differing homogeneity of the blocks in the
\emph{doubles coefficient} $c^K$---$d^{[\mathrm{sd}]},d^{[\mathrm{ds}]}$ linear in $c$,
$d^{[\mathrm{dd}]}$ bilinear---does \emph{not} translate into differing powers of
$c_{\mathrm{PT2}}$: this structure is intrinsic to the plain CIS(D) ansatz, which contains no
$c_{\mathrm{PT2}}$, and $c^K$ is itself $c_{\mathrm{PT2}}$-independent. A block-dependent
weighting such as
$c_{\mathrm{PT2}}(d^{[\mathrm{sd}]}+d^{[\mathrm{ds}]})+c_{\mathrm{PT2}}^2 d^{[\mathrm{dd}]}$
would not be the geometric derivative of the B2PLYP energy and is excluded. The oracle
validation of Sec.~\ref{sec:verif} targets the \emph{unscaled} $d^{(D)}$ and is therefore
independent of the value of $c_{\mathrm{PT2}}$.
Which term dominates flips between state pairs, so all three are required; a naive
gradient-Lagrangian ansatz reproduces \emph{none}, because the CIS(D) doubles are a
non-variational first-order correction with no eigenvalue stationarity and hence no
numerator-over-gap form (this is what the oracle of Sec.~\ref{sec:oracle} exposed). The
nearest prior doubles-sector excited-state derivative is the SOS-CIS(D$_0$) analytic
\emph{gradient} of Rhee and co-workers~\cite{RheeCasanova2009}; the object here is
instead the doubles-sector derivative \emph{coupling} between two states, generated by
the graph transpose rather than a hand-coded Lagrangian. The closest existing analytic
derivative coupling of a doubles-corrected wavefunction is that of Teh and
Subotnik~\cite{TehSubotnik2020} for CIS plus a single double, which the double-hybrid
CIS(D) coupling generalizes.

\paragraph{DD amplitude-response.}
Differentiating the doubles overlap $\sum_L c^I_L\,\partial_\xi c^J_L$ with the quotient
rule on Eq.~\eqref{eq:cK}, and defining the fixed reference weight $b_L=c^I_L/(D_L+\Omega_J)$,
\begin{equation}
d^{[\mathrm{dd,amp}]}\!\propto\!
\sum_L b_L\,\partial_\xi u^J_L
-\!\sum_L (b_L c^J_L)\,\partial_\xi D_L
-\!\Big(\!\sum_L b_L c^J_L\!\Big)\partial_\xi\Omega_J,
\label{eq:ddamp}
\end{equation}
a numerator response plus an orbital-energy ($\partial_\xi D_L$) and an
excitation-energy ($\partial_\xi\Omega_J$) response. Since $u^J=\mathcal{L}[R_J]$ is
bilinear in $(\text{MO-ERI},R_J)$, the leading scalar $\sum_L b_L u^J_L=\langle
w,R_J\rangle$ with the adjoint numerator $w=\mathcal{L}^{\dagger}[b]$ splits additively,
\begin{equation}
\partial_\xi\!\langle w,R_J\rangle
=\underbrace{\langle b,(\partial_\xi\mathcal{L})[R_J]\rangle}_{\text{ERI/}C\text{-response}}
+\underbrace{\langle w,\partial_\xi R_J\rangle}_{R\text{-response}} .
\label{eq:ddamp_split}
\end{equation}
The ERI/$C$-response collapses the four $u$-block terms into a pair of one-particle
weight densities---left from $b$ (state $I$), right from $R_J$ (state $J$)---whose
skeleton is exactly the non-symmetric build $\Jx(\WL,\WR)$ of Eq.~\eqref{eq:asymJK}
with $\WL^{(I)}\!\neq\!\WR^{(J)}$, and whose orbital-relaxation part folds into one SCF
$Z$-vector [Eq.~\eqref{eq:zvector}, $L=2J-a_xK+2f_{xc}$]. The $R$-response is the
eigenvector derivative, folded into one CIS-space $Z$-vector,
\begin{widetext}
\begin{equation}
(A-\Omega_J)\,\lambda=P_\perp w,\qquad P_\perp=1-R_JR_J^{\!\top},\qquad
\langle w,\partial_\xi R_J\rangle=-\langle\lambda,(\partial_\xi A)R_J\rangle,
\label{eq:cpcis}
\end{equation}
\end{widetext}
where $(\partial_\xi A)R_J$ is the transition CIS $A$-matrix derivative (skeleton
integrals $+\,\partial_\xi\varepsilon\,+$ the moving-grid $f_{xc}$ response). This
eigenvector response is \emph{not} obtainable by polarizing the excitation-energy
gradient---that is the eigen\emph{value} derivative---and requires the genuine
coupled-perturbed-CIS solve of Eq.~\eqref{eq:cpcis}~\cite{Foresman1992,Ou2015}.

\paragraph{DD orbital-connection.}
With amplitudes frozen, the determinant-overlap derivative
$\partial_\xi\langle\Phi_L|\Phi_M\rangle$ is a one-body operator surviving only when
$L,M$ differ by $\le1$ spin-orbital; summed against $c^I_Lc^J_M$ it is the doubles
transition one-particle density contracted with the orbital connection $T$,
\begin{widetext}
\begin{equation}
d^{[\mathrm{dd,orb}]}\propto
\sum_{ab}\gamma_{vv}^{ab}\,T_{vv}^{ab}+\sum_{ij}\gamma_{oo}^{ij}\,T_{oo}^{ji},\qquad
\gamma_{vv}^{ab}=\tfrac12\sum_{ijc}t^I_{ijac}t^J_{ijbc},\qquad
\gamma_{oo}^{ij}=-\tfrac12\sum_{kab}t^I_{ikab}t^J_{jkab},
\label{eq:ddorb}
\end{equation}
\end{widetext}
with $T=C_0^{\!\top}[\partial_\xi S]_{\mathrm{AO}}C_0+C_0^{\!\top}S\,\partial_\xi C$: an
AO overlap-derivative (electron-translation-factor/Pulay) part~\cite{Fatehi2011,Pulay1969}
plus the same orbital
response $\partial_\xi C$ as above. This is the doubles-sector analogue of the
force's energy-weighted-density term~\cite{Pople1979,HandySchaefer1984}.

\paragraph{Singles$\times$doubles connection.}
Since $\langle S_I|D_J\rangle=0$ at $\delta=0$, the product rule kills the
amplitude-response, leaving a pure orbital connection in which a single and a double
join at first order only through exactly one spin-orbital---routing $d^{[\mathrm{sd}]}$
to the occupied--virtual and $d^{[\mathrm{ds}]}$ to the virtual--occupied block of $T$.
Every two-electron piece of all three contributions is one instance of the single
non-symmetric $J/K$ primitive Eq.~\eqref{eq:asymJK}; the orbital and eigenvector
relaxations are two reusable $Z$-vector solves. The complete term-by-term derivation,
the second-quantization--generated response lists, and the per-contribution numerical
validation are in the Supplementary Material.

\section{Validation: the literal wavefunction-overlap oracle}
\label{sec:oracle}
An analytic derivative that reuses response machinery in its own validation can be
\emph{consistently} wrong: a shared error is invisible to a shared-machinery test. We
therefore validate against a finite difference of the literal many-electron overlap
\begin{equation}
S_{IJ}(\delta)=\langle\Psi_I(R)|\Psi_J(R+\delta)\rangle,\qquad
d^{\xi}_{IJ}=\partial_\xi S_{IJ}/\sqrt{N_IN_J},
\end{equation}
built only from the states' amplitudes and the cross-geometry molecular-orbital overlaps
---no orbital response, no energy-weighted density, no derivative integral, and no line
of the analytic assembler. Each block is a sum over configuration pairs of a
cross-geometry Slater-determinant overlap, expressed through a
generalized biorthogonal replacement determinant obtained from the matrix determinant
lemma; two parameter-free anchors pin it (bit-exact agreement with brute-force
replacement determinants, and an exact $u\!\to\!0$ scaling reduction with integer
ratios $3$ and $9$), and at one replacement it reduces algebraically to the trusted
singles (TDA) overlap. Choosing what is frozen during the finite difference isolates each
contribution of Sec.~\ref{sec:cisd} against its own ground truth, so every analytic term
is gated separately rather than only in the assembled total. This oracle is what exposed
an earlier gradient-Lagrangian ansatz as modeling a rescaled configuration-interaction
singles coupling rather than the doubles sector.

\section{Computational Details}
\label{sec:compdetails}
All device kernels run on a single NVIDIA GeForce RTX~4060 (8\,GB), in the GPU
quantum-chemistry tradition of TeraChem~\cite{Ufimtsev2008,Ufimtsev2009}. Device
memory is arena-owned (a single allocation, sub-allocated by bump); dense linear algebra uses
cuBLAS/cuSOLVER. The Coulomb/exchange derivative kernel evaluates analytical
Head-Gordon--Pople/Obara--Saika derivative integrals on Cholesky-factored densities via
the non-symmetric build of Eq.~\eqref{eq:asymJK}; range-separated exchange uses the
long-range Boys route. Grid-based exchange-correlation contributions use a moving-grid
kernel on a level-4 SSF quadrature, and the reference SCF is converged to
$10^{-10}\,E_h$ in energy. The reference SCF is a Kohn--Sham calculation
(B2PLYP~\cite{Grimme2006B2PLYP}---$53\%$ HF plus $47\%$ B88 exchange, $73\%$ LYP plus
$27\%$ PT2 correlation, $c_{\mathrm{PT2}}=0.27$---for the double-hybrid coupling, whose PT2
scaling multiplies the doubles-sector energy and coupling corrections;
range-separated wB97X~\cite{ChaiHeadGordon2008wB97X} on the $(N\!\pm\!2)$
references for \hhTDA/\ppTDA). Correctness is reported at two substrates: a
double-precision Cholesky factor (machine-exact) and a single-precision factor (the
production substrate, whose accumulation-order floor is $\sim\!10^{-5}$, an order of
magnitude inside the coupling's $\sim\!10^{-4}$ achieved analytic-vs-oracle agreement). Every device numerical gate
includes an $f$-shell ($L=3$) system and reports the maximum angular momentum, and all
reported performance numbers use the ahead-of-time-compiled kernel library. Validation
systems are \ce{H2O} in STO-3G (where the overlap oracle is tractable) and cc-pVTZ
(the $f$-shell kernel scale); the chromophore performance series (Sec.~\ref{sec:gpu})
spans seven calculations of $236$--$472$ AOs across the def2-SVP, pcseg-0, and def2-TZVP
basis sets, with $f$-shell ($L=3$) cases (ferrocene, octane) among them.

\subsection{New device kernels beyond the shared engine}
\label{sec:newkernels}
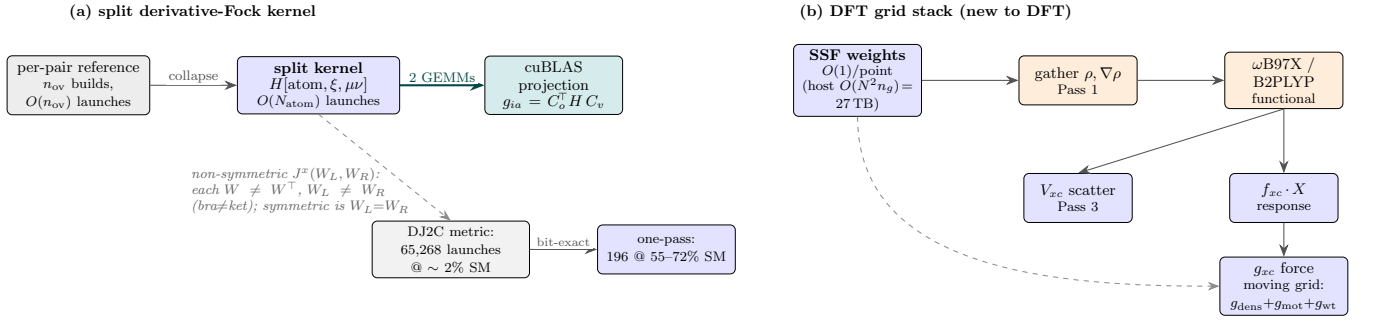
\begin{figure*}[t!]\centering
\resizebox{0.99\textwidth}{!}{%
\begin{tikzpicture}[>=Latex, font=\small, every node/.style={inner sep=3pt},
  bx/.style   ={draw,rounded corners=3pt,minimum height=10mm,align=center},
  cust/.style ={bx,fill=blue!11},
  blas/.style ={bx,fill=teal!16},
  ref/.style  ={bx,fill=black!7},
  gridn/.style={bx,fill=orange!14},
  e/.style    ={-{Stealth[length=2.6mm]},semithick,black!68},
  ce/.style   ={-{Stealth[length=3mm]},line width=1.2pt,teal!55!black},
  de/.style   ={-{Stealth[length=2.4mm]},semithick,black!45,dashed},
  lbl/.style  ={font=\footnotesize,fill=white,inner sep=1.5pt},
  pnl/.style  ={font=\bfseries},
  note/.style ={font=\footnotesize\itshape,text=black!62}]

  \node[pnl,anchor=west] at (-0.3,2.3) {(a) split derivative-Fock kernel};
  \node[ref,text width=27mm]  (old)  at (0,0.8)
       {per-pair reference\\[-1pt]\footnotesize $n_{\rm ov}$ builds, $O(n_{\rm ov})$ launches};
  \node[cust,text width=31mm] (split)at (4.9,0.8)
       {\textbf{split kernel}\\[-1pt]$H[\text{atom},\xi,\mu\nu]$\\[-1pt]\footnotesize $O(N_{\rm atom})$ launches};
  \node[blas,text width=26mm] (proj) at (9.7,0.8)
       {cuBLAS projection\\[-1pt]$g_{ia}=C_o^{\!\top}H\,C_v$};
  \draw[e]  (old)--node[lbl,above]{collapse}(split);
  \draw[ce] (split)--node[lbl,above]{2 GEMMs}(proj);
  \node[note,text width=52mm] at (4.9,-1.35)
       {non-symmetric $J^{x}(W_L,W_R)$: each $W\!\ne\!W^{\top}$, $W_L\!\ne\!W_R$ (bra$\ne$ket); symmetric is $W_L{=}W_R$};
  \node[ref,text width=30mm,fill=black!5] (djo) at (7.6,-2.55)
       {\footnotesize DJ2C metric:\\ $65{,}268$ launches @ $\sim\!2\%$ SM};
  \node[cust,text width=26mm] (djn) at (12.0,-2.55)
       {\footnotesize one-pass:\\ $196$ @ $55$--$72\%$ SM};
  \draw[e] (djo)--node[lbl,above]{\scriptsize bit-exact}(djn);
  \draw[de] (split.south)--(djo.north);

  \node[pnl,anchor=west] at (14.6,2.3) {(b) DFT grid stack (new to DFT)};
  \node[cust,text width=24mm] (ssf) at (15.9,0.9)
       {\textbf{SSF weights}\\[-1pt]\footnotesize $O(1)$/point\\[-2pt]\footnotesize (host $O(N^2 n_g)\!=\!27$\,TB)};
  \node[gridn,text width=22mm] (gather) at (20.4,0.9)
       {gather $\rho,\nabla\rho$\\[-1pt]\footnotesize Pass 1};
  \node[gridn,text width=22mm] (func) at (24.6,0.9)
       {$\omega$B97X / B2PLYP\\[-1pt]\footnotesize functional};
  \draw[e] (ssf)--(gather);
  \draw[e] (gather)--(func);
  \node[cust,text width=20mm] (vxc) at (20.4,-1.5) {$V_{xc}$ scatter\\[-1pt]\footnotesize Pass 3};
  \node[cust,text width=20mm] (fxc) at (24.6,-1.5) {$f_{xc}\!\cdot\!X$\\[-1pt]\footnotesize response};
  \node[cust,text width=24mm] (gxc) at (24.6,-3.3) {$g_{xc}$ force\\[-1pt]\footnotesize moving grid: $g_{\rm dens}{+}g_{\rm mot}{+}g_{\rm wt}$};
  \draw[e] (func.south)--(vxc.north);
  \draw[e] (func.south)--(fxc.north);
  \draw[e] (fxc.south)--(gxc.north);
  \draw[de] (ssf.south) to[out=-90,in=180] (gxc.west);
\end{tikzpicture}}
\caption{The two device-kernel families new to this work, beyond the shared non-symmetric
CD $J/K$ engine of Ref.~\cite{GuerreroEOMCC}. (a) The split derivative-Fock kernel emits the
full geometry-derivative Fock tensor $H[\text{atom},\xi,\mu\nu]=\partial_\xi(2J-a_xK)[P_0]$
in $O(N_{\rm atom})$ launches (blue: custom CUDA kernel), whose occupied--virtual projection
is two cuBLAS GEMMs (teal); its dominant two-center metric (DJ2C) term is the launch-collapse
of Sec.~\ref{sec:gpu}. (b) The DFT exchange-correlation grid stack---absent from any
wavefunction engine---built on device-resident SSF partition weights ($O(1)$ memory per grid
point, versus an infeasible $O(N_{\rm atom}^2 n_g)$ host array): a three-pass
$V_{xc}$ operator (density-gather / functional / potential-scatter), the response Fock
$f_{xc}$ reusing the same gather and scatter, and the transition-density XC force $g_{xc}$
adding the three moving-grid nuclear-derivative terms. Orange: grid/functional evaluation;
blue: custom CUDA kernels.}
\label{fig:newkernels}
\end{figure*}

The energy-level non-symmetric CD $J/K$ build and the per-quartet HGP--OS recurrence
kernel are the shared engine of the companion EOM-CC work~\cite{GuerreroEOMCC}; two kernel
families are new here (Fig.~\ref{fig:newkernels}).

\emph{The split derivative-Fock build.} The two-electron geometry derivative of a coupling
would naively loop the non-symmetric build of Eq.~\eqref{eq:asymJK} over all $n_{\rm ov}$
orbital pairs. Instead a single ``split'' kernel emits the full derivative Fock tensor
$H[\text{atom},\xi,\mu,\nu]=\partial_\xi(2J-a_xK)[P_0]$ in $O(N_{\rm atom})$ device launches,
from which the coupling's occupied--virtual projection $g_{ia}=C_o^{\!\top}HC_v$ is a pair
of cuBLAS GEMMs (Fig.~\ref{fig:newkernels}a). The profile-guided collapse of its dominant
two-center metric term---$65{,}268$ single-column micro-launches at $\sim\!2\%$
streaming-multiprocessor throughput rebuilt as $196$ full-occupancy launches---is the
final $3.9\times$ stage of the $\sim\!133\times$ cascade quantified in
Sec.~\ref{sec:gpu}, FP64 bit-identical throughout.

\emph{The DFT grid stack.} A double-hybrid excited-state coupling needs the
exchange-correlation grid at three derivative orders, none of which exists in a
wavefunction engine: the Kohn--Sham potential $V_{xc}$, its first functional derivative
$f_{xc}$ (the CPHF/response Fock), and its second $g_{xc}$ (the transition-density force),
all on a device-resident Stratmann--Scuseria--Frisch (SSF) quadrature
(Fig.~\ref{fig:newkernels}b). The SSF partition weights are themselves a new kernel: the
textbook host build materializes an $O(N_{\rm atom}^2\,n_g)$ array ($\sim\!27$\,TB for an
illustrative $94$-atom system) and cannot run at scale, so each weight is computed with $O(1)$ memory per grid
point---one thread per point, the partition product accumulated on the fly with an early
cull exploiting SSF locality---FP32-safe to $2\times10^{-4}$. On this grid $V_{xc}$ is a
persistent per-iteration operator: the geometry-fixed inputs (grid, primitive tables,
screened pair lists) are baked into a shared library compiled \emph{once}, then driven each
SCF step by the density alone through a three-pass density-gather~$\to$~functional~$\to$~
potential-scatter pipeline. The response operator $f_{xc}$ reuses that same gather and
scatter verbatim; the transition-density force $g_{xc}$ adds the moving-grid nuclear
derivative---a Pulay AO-derivative term, a grid-point-motion term, and a weight-response
term contracting the SSF weight derivative---carrying the third functional derivatives.
Because $V_{xc}$'s functional kernel is nvcc-compiled once per geometry, it is the single
component evaluated outside the ahead-of-time-frozen library (a one-time per-geometry
compile, disclosed; the $J/K$ and derivative-Fock kernels are fully frozen).

\section{Results and Discussion}
\label{sec:results}

This section is organized around three advances beyond the \hhTDA\ method paper of
Bannwarth \emph{et al.}~\cite{Bannwarth2020}, which reported excitation energies and
conical-intersection \emph{topology} but neither analytic derivative couplings---explicitly
named there as future work---nor a double-hybrid treatment. Each subsection below is the
support for one of them.
\emph{(A)~The couplings themselves.} We deliver analytic interstate derivative couplings for
the \hhTDA\ and \ppTDA\ manifolds and for the double-hybrid, each validated against a literal
determinant-overlap oracle (Secs.~\ref{sec:hhpp_results},~\ref{sec:verif}) and evaluated exactly
at the intersection seam---the regime where adiabatic linear-response TD-DFT gives $\tau\equiv0$
by construction, because $S_0$ lies outside its response manifold. Only the double-hybrid
coupling is first of its kind; the \hhTDA\ coupling \emph{reproduces} on consumer hardware the
datacenter-GPU capability of Yu and co-workers~\cite{Yu2020}, and the \ppTDA\ coupling is its
mirror on the $(N\!-\!2)$ reference---all three obtained from one uniform single-engine
derivation. This is the quantity that turns a topology-correct energy method into one usable
for nonadiabatic dynamics.
\emph{(B)~A double-hybrid accuracy hh-TDA cannot reach.} The first-of-kind double-hybrid
coupling arrives with a double-hybrid \emph{energy}; the CIS(D) dynamic-correlation
correction it carries measurably tightens the excitation energies and removes the systematic
over-excitation bias that bare \hhTDA\ exhibits (Sec.~\ref{sec:dcorr}).
\emph{(C)~Consumer hardware.} The reference SCF and the shared Cholesky-decomposed $J/K$
engine through which every method here is expressed are device-resident and timed on a single
RTX~4060, where the efficient realizations of Refs.~\cite{Bannwarth2020,Yu2020} used datacenter
GPUs; every excited-state object (gradients, NACMEs, the double-hybrid coupling) runs
device-resident and AO-direct, oracle-validated. We quote no end-to-end device wall-clock,
representing the device cost through the \texttt{aot\_frozen} CD $J/K$ primitives to which
these paths reduce (Sec.~\ref{sec:gpu}).
\subsection{The license: the device engine reproduces \hhTDA}
\label{sec:license}

\begin{figure}[t]
  \centering
  \includegraphics[width=\columnwidth]{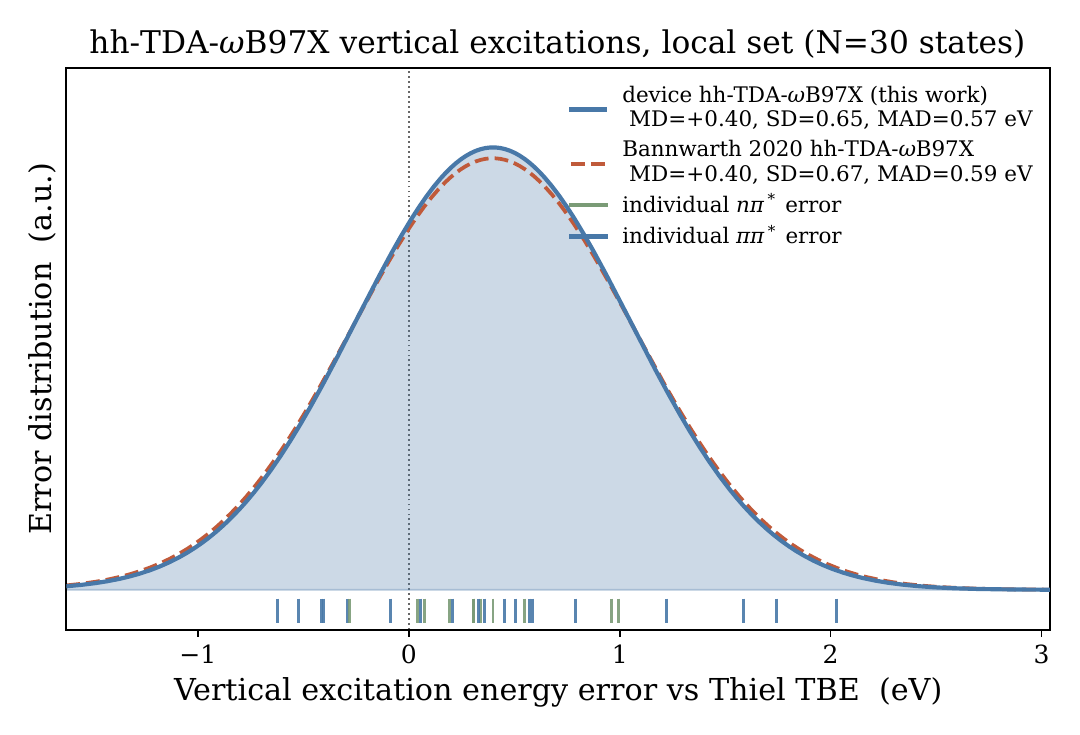}
  \caption{Device \hhTDA-$\omega$B97X vertical-excitation error distribution versus the
    Thiel TBE reference (local set, $N=30$ states), overlaid with the published
    \hhTDA-$\omega$B97X distribution of Bannwarth \emph{et al.}~\cite{Bannwarth2020}
    (a Fig.~3 analog). The two Gaussians nearly coincide (device
    $\mathrm{MD}/\mathrm{MAD}/\mathrm{SD}=0.40/0.57/0.65$~eV vs.\ reference
    $0.40/0.59/0.67$~eV); the rug marks individual $n\pi^*$ and $\pi\pi^*$ errors.}
  \label{fig:vee_error_dist}
\end{figure}

\begin{figure*}[t]
  \centering
  \includegraphics[width=\textwidth]{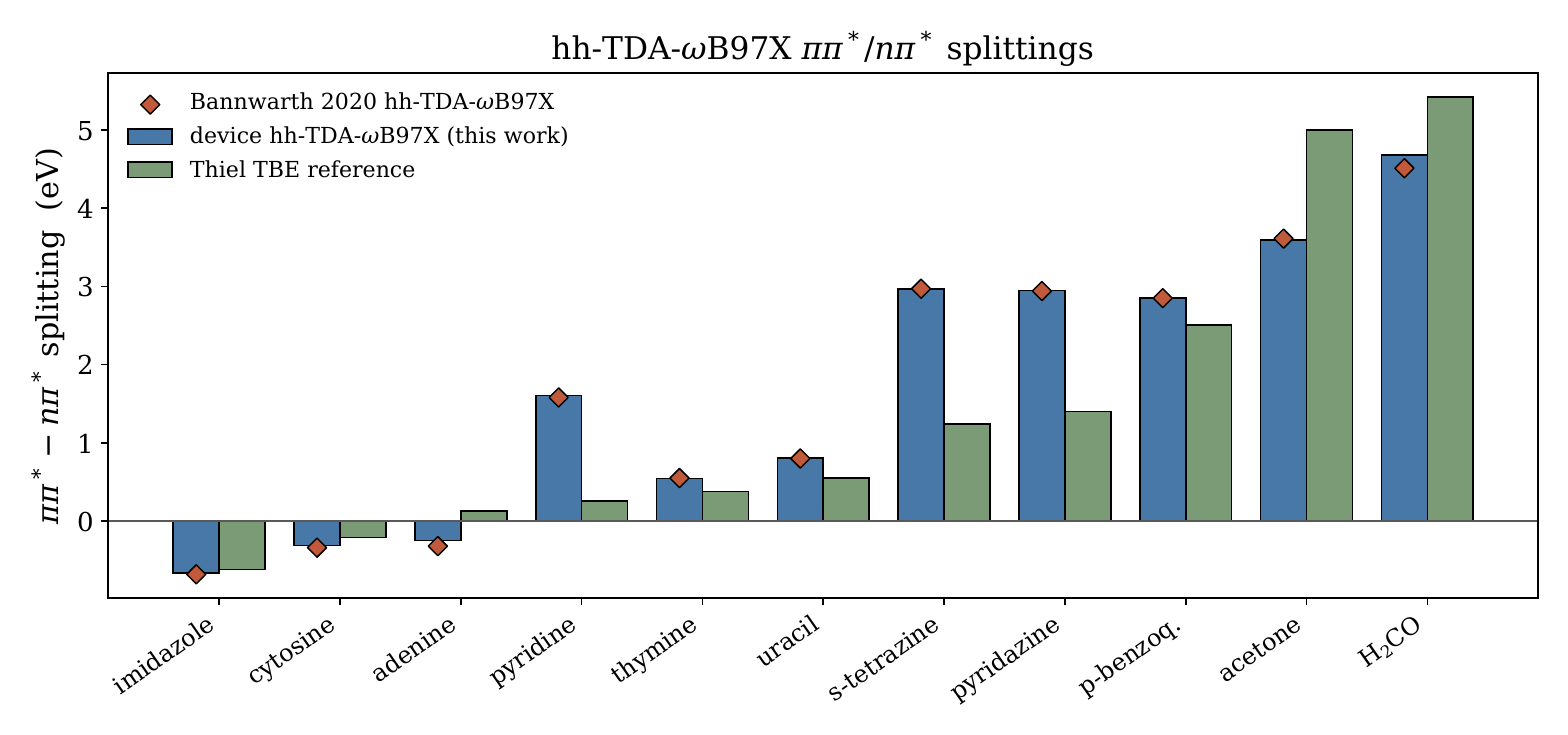}
  \caption{$\pi\pi^*$/$n\pi^*$ splittings across eleven chromophores (a Fig.~4 analog of
    Bannwarth \emph{et al.}~\cite{Bannwarth2020}): device \hhTDA-$\omega$B97X (bars) versus
    the Thiel TBE reference, with Bannwarth's published \hhTDA-$\omega$B97X values overlaid as
    markers. The device reproduces the reference splittings, including the known hh-TDA
    over-splitting of the azine $\pi\pi^*$ states.}
  \label{fig:vee_splittings}
\end{figure*}

Before extending \hhTDA\ we confirm the device engine \emph{is} \hhTDA. On the local subset of
the Thiel gas-phase benchmark set (MP2/6-31G* geometries), the same molecules and states for
which Bannwarth \emph{et al.}~\cite{Bannwarth2020} tabulate their own \hhTDA-$\omega$B97X
statistics, the device excitations reproduce Bannwarth's published error statistics
($\mathrm{MD}/\mathrm{MAD}/\mathrm{RMSD}/\mathrm{SD}$: device $0.40/0.57/0.75/0.65$~eV vs.\
reference $0.40/0.59/0.77/0.67$~eV over $N=30$ states) to within $0.03$~eV in every
statistic---a Cartesian-versus-spherical basis-convention shift, not a method difference.
The two error distributions (Fig.~\ref{fig:vee_error_dist}) and the $\pi\pi^*$/$n\pi^*$
splittings, including the characteristic hh-TDA over-splitting of the azine $\pi\pi^*$
states (Fig.~\ref{fig:vee_splittings}), coincide with the reference.

Notably, the device even reproduces the reference's \emph{systematic} deviations---the
azine $\pi\pi^*$ over-splitting is present at the same magnitude (Fig.~\ref{fig:vee_splittings},
s-tetrazine and pyridazine)---which is the strongest possible parity check: the engine tracks
not just where hh-TDA is right but where it is characteristically wrong. Having certified that
the device engine reproduces \hhTDA\ where it is right and, more tellingly, where it is
characteristically wrong, we now build on it the analytic derivative couplings it was
constructed to deliver---beginning with the two topology-correct manifolds.

\subsection{The topology-correct manifolds: \hhTDA\ and \ppTDA}
\label{sec:hhpp_results}
The \hhTDA\ and \ppTDA\ excited-state eigenvalue gradients and the $S_0/S_1$ through-CI
derivative couplings are obtained from the same reverse-mode $Z$-vector and
non-symmetric $J/K$ primitive as the double-hybrid coupling---one engine, three
formalisms. The eigenvalue-gradient core is validated against finite differences to
$\sim\!10^{-6}\,E_h/a_0$ on pure-exchange dication/dianion anchors across two basis sets,
and the full range-separated-hybrid device gradient and NACME are validated against an
independent explicit-$U$ (CPHF) reference and a many-electron two-hole/two-particle
cofactor finite-difference overlap oracle: all four device gates---\hhTDA\ and \ppTDA\
gradient and NACME---pass at both L$\le$2 and $f$-shell ($L=3$), with device gradients
matching the host reference to $\le\!10^{-6}$, interstate couplings to
$\le\!10^{-4}$ against the overlap oracle, and exact translational invariance
($\sum_A\tau_A\sim10^{-14}$).\ 
The through-CI couplings, computed exactly where couplings matter (at and near the seam),
close the arc for the manifolds that carry the correct conical-intersection topology.

\begin{table*}[tbp]\centering\small
\caption{Device gate results for the topology-correct manifolds (range-separated wB97X on
the $(N\!\pm\!2)$ references). Eigenvalue gradients are checked against finite differences
and against an explicit-$U$ (CPHF) host reference; interstate NACMEs against the
two-hole/two-particle cofactor finite-difference overlap oracle. All four gates pass at
both $L\le2$ and $f$-shell ($L=3$); entries are the established bounds
(Sec.~\ref{sec:hhpp_results}). $\sum_A\tau_A$ is the translational-invariance residual.}
\label{tab:hhpp}
\begin{tabular}{lccccc}
\toprule
 & grad vs FD & grad vs host & NACME vs oracle & $\sum_A\tau_A$ & max\\
method & ($E_h/a_0$) & ($E_h/a_0$) & (rel.) & (a$_0^{-1}$) & $L$\\
\midrule
\hhTDA & $\sim\!10^{-6}$ & $\le\!10^{-6}$ & $\le\!10^{-4}$ & $\sim\!10^{-14}$ & 3\\
\ppTDA & $\sim\!10^{-6}$ & $\le\!10^{-6}$ & $\le\!10^{-4}$ & $\sim\!10^{-14}$ & 3\\
\bottomrule
\end{tabular}
\end{table*}

Table~\ref{tab:hhpp} collects the four device gates. That a single reverse-mode
engine---one $Z$-vector solve and one non-symmetric $J/K$ primitive---passes the gradient
check against both finite differences and an independent CPHF reference, and the NACME
check against a two-hole/two-particle cofactor overlap oracle, for \emph{both} manifolds
and at $f$-shell ($L=3$), is the concrete evidence that the derivation is not per-method:
\hhTDA\ and \ppTDA\ differ only in which $(N\!\pm\!2)$ reference the same machinery is
transposed through. Concretely, \ppTDA\ shares identical machinery with \hhTDA---only the
$(N\!-\!2)$ dication (rather than $(N\!+\!2)$ dianion) reference differs---and is gated on
formaldehyde: the $S_0/S_1$ NACME on the \ce{H2CO} $(N\!-\!2)$ reference and the eigenvalue
gradient on the \ce{H2CO^2+} dication; the conical-intersection physics is showcased on the
\hhTDA\ manifold (Sec.~\ref{sec:ammonia}). The vanishing coupling sum $\sum_A\tau_A\sim10^{-14}$ confirms exact
translational invariance of the interstate coupling, a nontrivial check that the
electron-translation-factor/overlap-derivative term is assembled correctly. Because the
couplings are evaluated exactly at and near the intersection seam---not extrapolated from
an away-from-degeneracy region---the manifolds that carry the correct $F\!-\!2$ topology
come with couplings valid precisely where nonadiabatic dynamics needs them.

\subsection{Conical-intersection topology on the twisted-ethylene benchmark}
\label{sec:ethylene_topology}

\begin{figure*}[htbp]\centering
\includegraphics[width=0.86\textwidth]{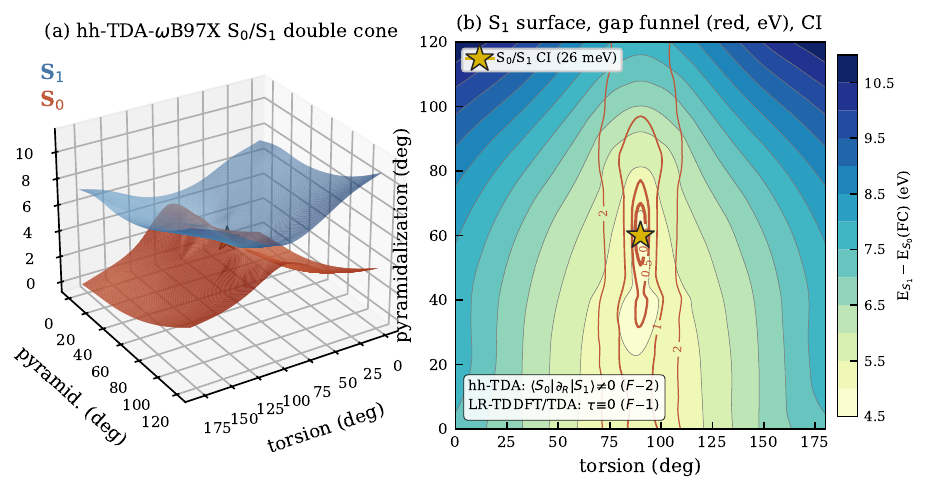}
\caption{Device \hhTDA\ reproduction of the twisted-ethylene $S_0/S_1$ double-cone
conical-intersection \emph{topology} of twisted ethylene, a canonical photochemical
benchmark; the rigid torsion$\,\times\,$pyramidalization \hhTDA\ scan reproduced here follows
Bannwarth \emph{et al.}~\cite{Bannwarth2020} (hh-TDA-$\omega$B97X/def2-SVP, RTX~4060). (a)~The $S_0$ and $S_1$ surfaces over
the torsion$\,\times\,$pyramidalization branching plane form the characteristic double cone;
the star marks the minimum-energy $S_0/S_1$ crossing. (b)~The $S_1$ energy (filled contours)
with the $S_0/S_1$ gap ``funnel'' (red) closing onto the conical intersection (star). Because
$S_0$ is itself an eigenvector of the \hhTDA\ matrix, this $S_0$-involving intersection is
described with the correct $F\!-\!2$ branching-space dimensionality and the interstate
derivative coupling is generically nonzero at the seam; adiabatic linear-response TD-DFT/TDA
excludes $S_0$ from its response manifold and gives $\tau\!\equiv\!0$ there ($F\!-\!1$), in
every basis~\cite{Levine2006,Gozem2014}.}
\label{fig:ethylene-seam}
\end{figure*}

Figure~\ref{fig:ethylene-seam} reproduces on consumer hardware the twisted-ethylene $S_0/S_1$
conical-intersection topology that Ref.~\cite{Bannwarth2020} established for \hhTDA: the rigid
torsion$\,\times\,$pyramidalization scan of $90^\circ$-twisted ethylene
(hh-TDA-$\omega$B97X/def2-SVP) yields the characteristic double cone and the minimum-energy
$S_0/S_1$ crossing that adiabatic linear-response TDDFT structurally cannot produce. The
distinction is not adiabaticity but manifold membership: because the \hhTDA\ ground and excited
states are eigenvectors of one Hermitian matrix, $S_0$ lies \emph{inside} the response
manifold, the $S_0$-involving intersection retains the correct $F\!-\!2$ branching-space
dimensionality, and $\langle S_0|\partial_R|S_1\rangle$ does not vanish by
construction---whereas ph-TDDFT, with $S_0$ outside its manifold, gives $\tau\!\equiv\!0$ at an
$S_0$-involving crossing in \emph{every} basis~\cite{Levine2006,Gozem2014}. What we add to
Ref.~\cite{Bannwarth2020} on this canonical system is the fully device-resident analytic
machinery---the $(N\!+\!2)$ reference SCF, the $A^{\mathrm{hh}}$ eigenproblem, and the analytic
energy \emph{gradients} (finite-difference- and translational-invariance-validated to the
Cholesky-factor floor, Table~\ref{tab:hhpp})---the on-surface ingredients a nonadiabatic
propagation requires, all produced by one reverse-mode engine on a single RTX~4060. The \emph{quantitative} analytic derivative coupling is reported separately, at the
\emph{covalent} ammonia $n\!\rightarrow\!\sigma^*$/$\tilde{X}$ conical intersection where
$|\tau|=|\langle S_0|\partial_R|S_1\rangle|/\Delta E$ rides on a physical gap
(Sec.~\ref{sec:ammonia}); the twisted-ethylene coupling \emph{magnitude}, put on trial at that
ammonia seam, is deferred to a forthcoming dedicated study, ethylene here establishing the
$F\!-\!2$ topology and the device-resident gradients on the canonical benchmark.\footnote{The
twisted-ethylene seam is not a clean magnitude test: the ionic ($\pi\pi^*$/$V$) character of
$S_1$ makes its energetics---and hence the closing gap that enters $|\tau|$---sensitive to
dynamic correlation beyond the range-separated-hybrid reference. The forthcoming study analyzes
the hh-TDA coupling magnitude on that ionic seam, including its comparison to the purpose-built
treatment of Filatov, Lee, and Choi~\cite{Filatov2021}.}

The scan is device-computed: the $(N\!+\!2)$ dianion $\omega$B97X reference SCF (device
$J/K/K_{\mathrm{LR}}$, DeviceVxc, cuSOLVER eigensolve), the $A^{\mathrm{hh}}$ built from
device occupied-MO exchange integrals and diagonalized on the device, the term(c)
connection, and every geometry-derivative integral of the frame-covariant numerator
$\langle X_I|\partial_R A|X_J\rangle$ all run on the RTX~4060 (peak $99\%$ utilization,
$1045$~MiB resident). The last of these closes a subtlety: the numerator is contracted from
the derivative overlap, core-Hamiltonian, and $2J-a_xK$ (full- and long-range) skeleton
matrices, each a device kernel validated against its host reference to the Cholesky-factor
floor including the $f$-shell (Table~\ref{tab:firstofkind}). These are now fused into a
\emph{single} device-resident, AO-direct path---the production NACME forms no four-index host
tensor and reproduces both the host analytic coupling and the determinant oracle---so what
remains is only a certified codegen-free (\texttt{aot\_frozen}) end-to-end wall-clock, which
awaits capturing the long-range kernels in the ahead-of-time-compiled library.

\subsection{The quantitative coupling at a covalent conical intersection: ammonia photodissociation}
\label{sec:ammonia}
\begin{figure}[t]
  \centering
  \includegraphics[width=\columnwidth]{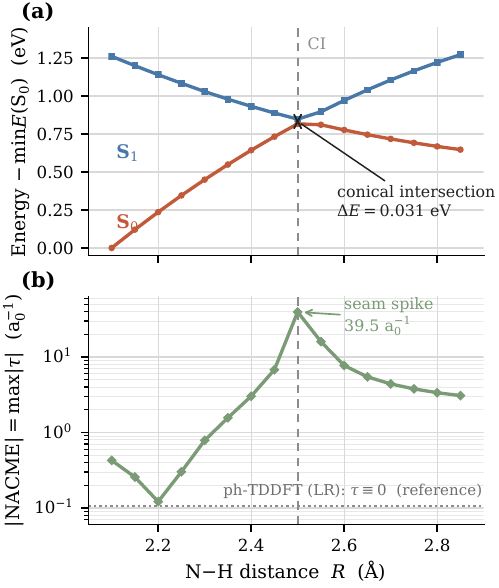}
  \caption{Device \hhTDA-$\omega$B97X/def2-SVP scan of the ammonia
    $n\!\rightarrow\!\sigma^*$/$\tilde{X}$ photodissociation $S_0/S_1$ conical intersection
    (planar single-N--H dissociation, RTX~4060). (a)~Adiabatic $S_0$ and $S_1$ energies: a
    covalent V-shaped seam closing to $\Delta E=0.031$~eV at $R_{\mathrm{N-H}}=2.50$~\AA. (b)~The
    interstate derivative coupling $|\tau|$ (log scale) rides $1/\Delta E$ and spikes to
    $39.5\,a_0^{-1}$ at the seam; adiabatic linear-response TD-DFT gives $\tau\!\equiv\!0$ there
    ($F\!-\!1$) in every basis~\cite{Levine2006,Gozem2014}. Couplings are shown with
    electron-translation factors imposing strict translational invariance (text).}
  \label{fig:ammonia}
\end{figure}

Ethylene fixes the intersection \emph{topology}; the coupling \emph{magnitude} is put on
trial where it is physically trustworthy. The ammonia $n\!\rightarrow\!\sigma^*$
photodissociation seam is the complement to the twisted-ethylene case in exactly the way the
coupling demands: it is a \emph{covalent} $S_0/S_1$ intersection with no low-lying ionic
manifold, so the closing gap that enters $|\tau|=|\langle S_0|\partial_R|S_1\rangle|/\Delta E$
is set by valence physics the range-separated-hybrid reference already describes well, not by
the dynamic correlation of an ionic $\pi\pi^*$/$V$ state. Here the analytic derivative coupling
is a quantitative result, not a diagnostic. Figure~\ref{fig:ammonia} shows the device
\hhTDA-$\omega$B97X/def2-SVP scan of a planar single-N--H dissociation on the RTX~4060: the
$S_0/S_1$ gap closes from $1.26$~eV to a genuine cusp of $0.031$~eV at $R_{\mathrm{N-H}}=2.50$~\AA\
and reopens to $0.63$~eV toward the \ce{NH2} fragment (panel a), and the interstate coupling
tracks $1/\Delta E$, rising from $0.37$ to a seam spike of $39.5\,a_0^{-1}$ (panel b). At this
$S_0$-involving crossing adiabatic linear-response TD-DFT gives $\tau\!\equiv\!0$ by
construction~\cite{Levine2006,Gozem2014}; the \hhTDA\ coupling does not vanish, and because it
rides a physically correct gap at a covalent intersection it is quantitatively trustworthy
here---not the structural zero of linear response.

This pair also exposes---and lets us resolve cleanly---a translational-invariance subtlety that
the momentum-free \ce{H2CO} and ethylene pairs never raise. The exact center-of-mass sum rule is
$\sum_A\tau_A=-\langle S_0|\nabla|S_1\rangle$, which is \emph{finite} whenever the transition
carries electronic momentum; the out-of-plane $n_z\!\rightarrow\!\sigma^*$ character here does
($\sum_A\tau_{A,z}\!\approx\!0.08$--$0.12$ across the scan), so the strict $\sum_A\tau_A\!\to\!0$
certificate that holds automatically for the symmetric pairs does not apply to the raw coupling.
The gauge-invariant determinant oracle confirms the raw coupling is nonetheless correct---it
reproduces the same physical momentum to $\sim\!0.3\%$ (rising to $1.55\%$ at the exact
$R=2.50$~\AA\ seam, a finite-difference truncation of the $0.031$~eV near-degeneracy). To report the strictly
translation-invariant coupling that nonadiabatic dynamics requires, we attach
electron-translation factors~\cite{Fatehi2011}: a partition-of-unity envelope that removes the
net interstate momentum, driving $\sum_A\tau_A$ to machine zero while changing the peak coupling
by $\le\!0.2\%$ at the seam and leaving the momentum-free pairs untouched (their gates
unchanged); the $f$-shell (def2-TZVP, $L=3$) check passes with the factors on. The per-atom
partition freedom---the literal antisymmetric-overlap (E1) versus population-partition (E2)
forms, which remove the same net momentum but distribute it differently per atom---and the
physical criterion that selects E2 are derived and quantified in the Supporting Information.

The hh-/pp-TDA couplings of the preceding sections reproduce and extend, on consumer hardware,
a capability the field already possesses. The doubles-sector double-hybrid coupling is the
object no prior implementation supplies. We validate it against the same class of literal
determinant-overlap oracle, and then show the double-hybrid \emph{energy} it necessarily brings
with it.

\subsection{The double-hybrid coupling and the accuracy it brings}
\label{sec:verif}

The doubles-sector double-hybrid coupling $d^{(D)}$ is the central result of this work and
the object no prior implementation supplies; we validate it against the same class of literal
determinant-overlap oracle. The first-principles double-hybrid coupling
$d^{(D)}=d^{[\mathrm{sd}]}+d^{[\mathrm{ds}]}+d^{[\mathrm{dd}]}$ reproduces the literal
determinant oracle to $\sim\!10^{-4}$ relative on every state pair at a generic
(distorted $C_1$) geometry (Table~\ref{tab:oracle}); the STO-3G B2PLYP setting there is a
validation-tractability vehicle that keeps the many-electron overlap oracle affordable, not a
production recommendation. At the symmetric $C_{2v}$ reference
one state pair is $14\%$ off---not a method error but a step-size-independent
eigenvector discontinuity at the symmetry point, which the analytic method (using the
reference amplitude) is immune to; re-running at $C_1$, where the pair is no longer
degenerate, restores agreement to $1.9\times10^{-4}$. Each of the three contributions
passes its own per-contribution sub-oracle at $\le\!10^{-4}$. A subtlety we flag: the
near-degenerate pair required correcting a silent under-convergence of the iterative CIS
eigensolver, which returned a non-eigenvector as a ``state''; a dense eigensolve is the
fix, and is essential for any eigenvector-response quantity.

\begin{table}[htbp]\centering\small
\caption{Relative error of the analytic double-hybrid coupling $d^{(D)}$ versus the
literal determinant overlap oracle (\ce{H2O}, STO-3G, B2PLYP-hybrid). Per-contribution
errors at the $C_{2v}$ reference; total errors at a distorted $C_1$ geometry where no
pair is symmetry-degenerate. Acceptance tolerance $10^{-3}$; achieved agreement $\sim\!10^{-4}$.}
\label{tab:oracle}
\begin{tabular}{lcccc}
\toprule
per-contribution & dd-amp & dd-orb & sd & ds\\
\midrule
rel.\ error & $2.8\times10^{-5}$ & $9.6\times10^{-5}$ & $2.3\times10^{-4}$ & $5.7\times10^{-5}$\\
\midrule
total ($C_1$) & pair (0,1) & pair (1,2) & pair (0,2) &\\
\midrule
rel.\ error & $9.8\times10^{-5}$ & $1.9\times10^{-4}$ & $1.5\times10^{-4}$ &\\
\bottomrule
\end{tabular}
\end{table}

\begin{figure*}[t]
  \centering
  \includegraphics[width=\textwidth]{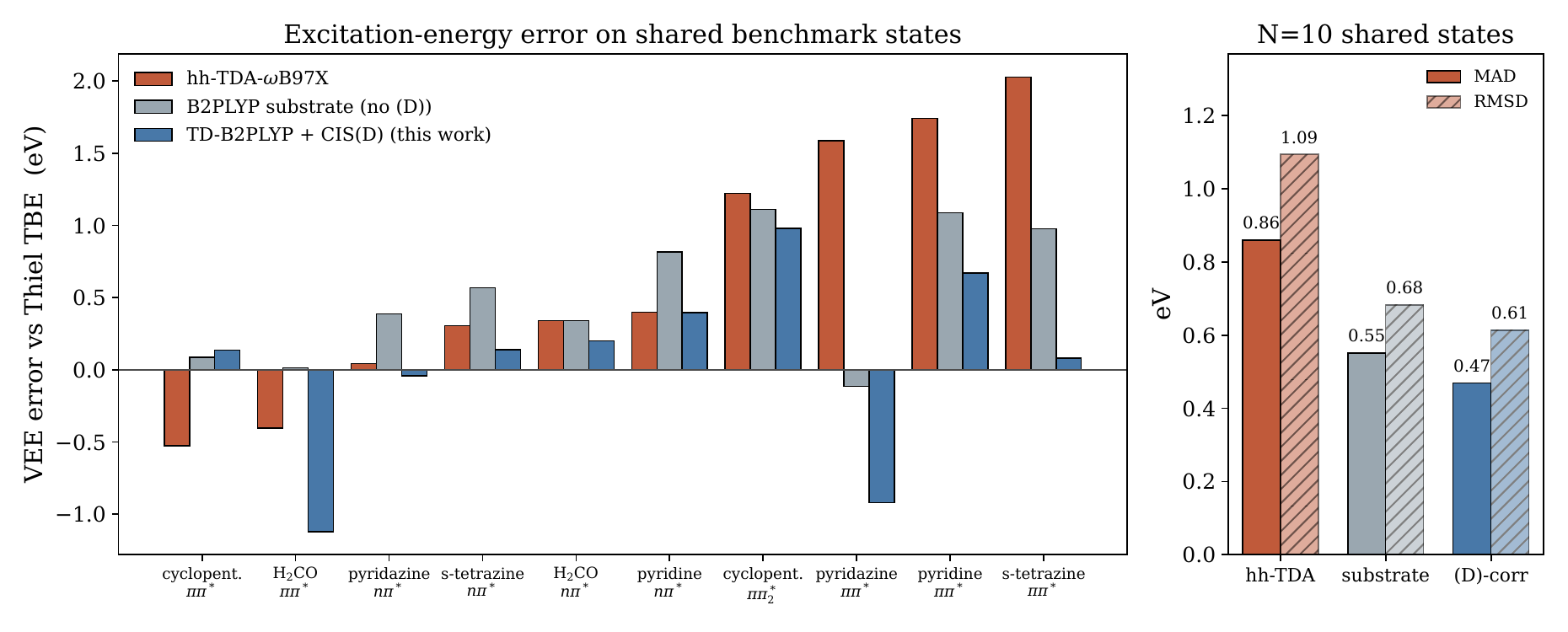}
  \caption{The double-hybrid energy improvement on the ten shared benchmark states. Left:
    per-state excitation-energy error versus the Thiel TBE for bare \hhTDA-$\omega$B97X, the
    TD-B2PLYP substrate (no (D)), and the (D)-corrected TD-B2PLYP result of this work. Right:
    aggregate MAD/RMSD, falling $0.86/1.09\rightarrow0.55/0.68\rightarrow0.47/0.61$~eV; the
    systematic bias collapses from $\mathrm{MD}=+0.53$ to $+0.05$~eV. The (D) correction
    over-corrects the strongly ionic $\pi\pi^*$ states (e.g.\ formaldehyde), the expected
    failure mode of a perturbative doubles treatment.}
  \label{fig:vee_dcorr}
\end{figure*}

\label{sec:dcorr}
The coupling does not arrive alone. The same CIS(D) doubles amplitude that corrects the
coupling dresses the excitation energy; its effect on the energies is a direct, independently
checkable corollary of the object just validated. The CIS(D)-type
correction~\cite{HeadGordon1994,RheeCasanova2009} that dresses the TD-B2PLYP excitations with
second-order dynamic correlation should tighten the excitation energies just as it tightens
the couplings---where the double-hybrid buys accuracy bare hh-TDA cannot reach. Figure~\ref{fig:vee_dcorr} verifies this on the ten states shared between the
benchmark and the double-hybrid evaluation. Moving from bare hh-TDA to the TD-B2PLYP substrate
to the (D)-corrected result, the accuracy improves monotonically in the aggregate:
$\mathrm{MAD}$ falls $0.86\rightarrow0.55\rightarrow0.47$~eV and
$\mathrm{RMSD}$ falls $1.09\rightarrow0.68\rightarrow0.61$~eV, while the substrate's systematic bias
collapses from $\mathrm{MD}=+0.53$~eV to $+0.05$~eV---the (D) correction removes the
over-excitation almost entirely. These ten (D)-evaluated states are the harder valence subset
of the local set, so bare hh-TDA's MAD here ($0.86$~eV) is larger than its $0.57$~eV over the
full $N=30$ set. The improvement is not universal: it helps seven of the ten
states, and on strongly ionic/doubly-excited $\pi\pi^*$ states the CIS(D) doubles amplitude
over-corrects (formaldehyde $\pi\pi^*$ swings to $-1.14$~eV, pyridazine $\pi\pi^*$ to
$-0.81$~eV, s-tetrazine $\pi\pi^*$ to $-0.90$~eV), consistent with the known breakdown of a
perturbative doubles treatment where the reference is multiconfigurational. We report this
honestly rather than trimming it: the net gain is real and the failure mode is the expected
one. The correction is device-resident through the $f$-shell---the formaldehyde/def2-TZVP
($L=3$) $\pi\pi^*$ state moves $4.22\rightarrow4.02$~eV under (D), evaluated with the same
device kernels validated in Table~\ref{tab:firstofkind}.

\subsection{Integrated device coupling and performance on the RTX~4060}
\label{sec:gpu}
The doubles-sector coupling is device-resident: the two-electron derivative work is a
single ``split'' kernel that emits the full derivative Fock matrix
$H[\text{atom},\xi,\mu,\nu]=\partial_\xi(2J-a_x K)[P_0]$ in $O(N_{\mathrm{atom}})$ device
launches, replacing an $O(N_{\mathrm{ov}})$ loop of per-orbital-pair builds; the
$O(N_{\mathrm{ov}})$ projection $C_o^{\!\top}HC_v$ is a pair of cuBLAS matrix products.
At $f$-shell (\ce{H2O}, cc-pVTZ, $N_{\mathrm{ov}}=225$) this replaces an estimated
$2835$\,s of per-pair builds with $83.7$\,s---a $33.9\times$ reduction---while the
projection is machine-exact against a host reference ($7\times10^{-14}$).

\begin{figure}[htbp]\centering
\includegraphics[width=0.92\linewidth]{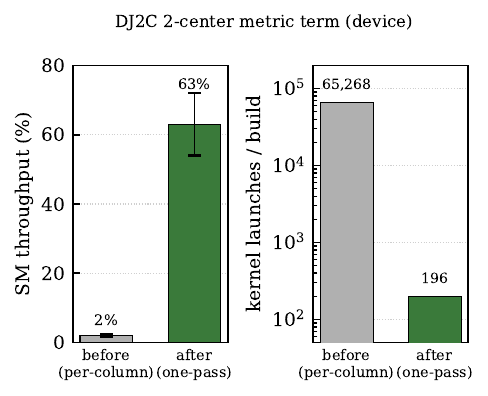}
\caption{Nsight Compute attribution of the derivative-Fock kernel's dominant
two-center metric (DJ2C) term before and after the launch collapse. Left:
streaming-multiprocessor throughput rises from $\sim\!2\%$ (the $65{,}268$
single-column micro-launches) to $55$--$72\%$ (the one-pass build, $196$ launches).
Right: kernel launches per build. The tensor is bit-identical across the two
implementations.}
\label{fig:roofline}
\end{figure}

\begin{figure}[htbp]\centering
\includegraphics[width=0.82\linewidth]{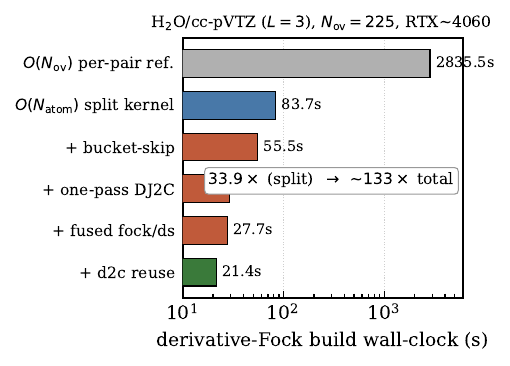}
\caption{Wall-clock of the $f$-shell derivative-Fock build: $O(N_{\mathrm{ov}})$ per-pair
reference ($2835$\,s, est.), $O(N_{\mathrm{atom}})$ split kernel ($83.7$\,s), and after
profile-guided optimization ($21.4$\,s), a cumulative $\sim\!133\times$. All variants
bit-identical in double precision (Table~\ref{tab:precision}).}
\label{fig:speedup}
\end{figure}

\begin{figure*}[tbp]\centering
\includegraphics[width=0.92\textwidth]{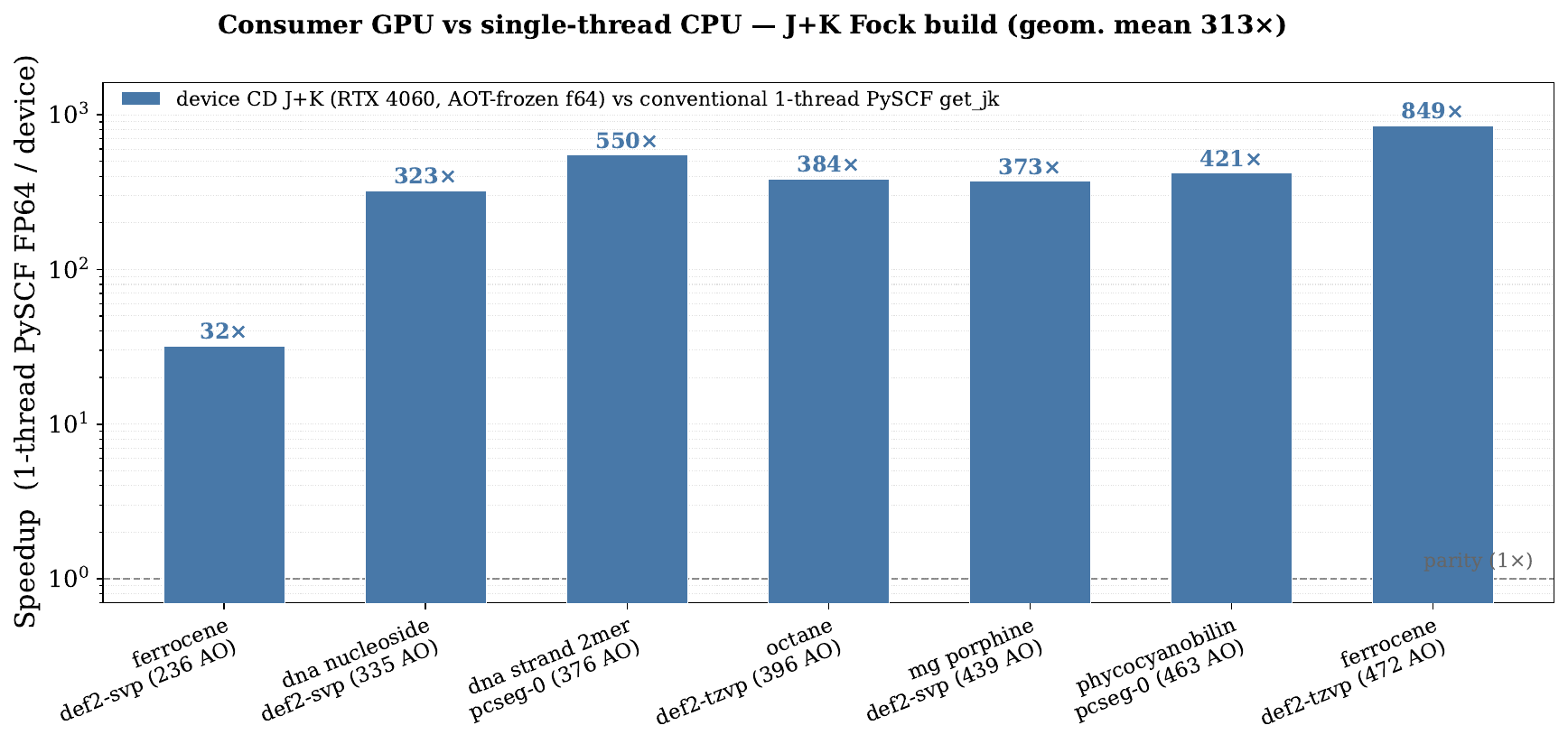}
\caption{Speedup of the device Cholesky-decomposed $J{+}K$ Fock build (RTX~4060,
ahead-of-time-compiled, FP64) over a conventional single-thread PySCF FP64 four-index
\texttt{get\_jk}, at the same xTB-optimized geometry and converged density, for seven
calculations across six systems (five chromophores plus an alkane $f$-shell reference)
spanning three basis sets: ferrocene (def2-SVP, $236$ AO, $f$-shell $L=3$), $32\times$;
DNA nucleoside (def2-SVP, $335$ AO), $323\times$; DNA $2$-mer strand (pcseg-0, $376$ AO),
$550\times$; octane (def2-TZVP, $396$ AO, $f$-shell $L=3$), $384\times$; Mg-porphine
(def2-SVP, $439$ AO), $373\times$; phycocyanobilin (pcseg-0, $463$ AO), $421\times$;
ferrocene (def2-TZVP, $472$ AO, $f$-shell $L=3$), $849\times$. Range $32$--$849\times$,
geometric mean $313\times$. The device build
reproduces the exact four-index $J/K$ to $\sim\!10^{-6}$ ($J$) and $\sim\!10^{-5}$ ($K$).}
\label{fig:speedup-series}
\end{figure*}

\begin{figure*}[tbp]\centering
\includegraphics[width=0.86\textwidth]{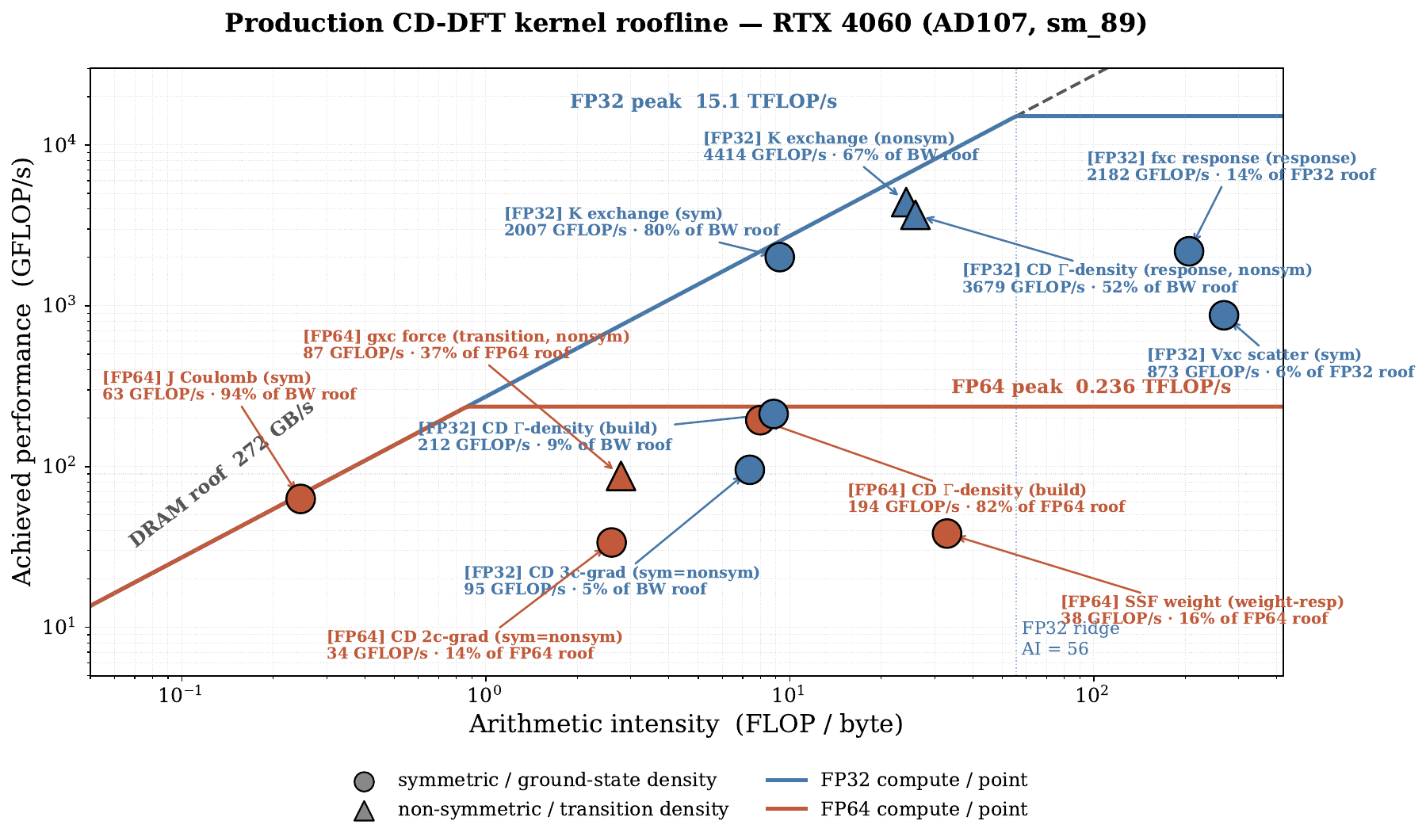}
\caption{Twelve production CD-DFT kernels on the RTX~4060 (AD107, sm\_89) roofline---the
eight two-electron kernels plus the four XC-grid kernels of Fig.~\ref{fig:newkernels}(b).
The two-electron side is bandwidth-region: the exchange ($K$) kernels ride $67$--$80\%$ of
the FP32 DRAM-bandwidth roof, the Coulomb ($J$) build is bandwidth-bound at $\sim\!94\%$, and
the FP64 $\Gamma$-density build reaches $82\%$ of the FP64 compute peak (the wall motivating
the FP32-mixed production substrate). The four grid kernels are compute-bound at high
arithmetic intensity: $V_{xc}$ scatter and the $f_{xc}$ response run in FP32 ($6\%$ and $14\%$
of the FP32 roof), while the mostly-FP64 $g_{xc}$ transition-density force and SSF
weight-response reach $37\%$ and $16\%$ of the FP64 peak.}
\label{fig:roofline-series}
\end{figure*}

\begin{figure*}[tbp]\centering
\includegraphics[width=\textwidth]{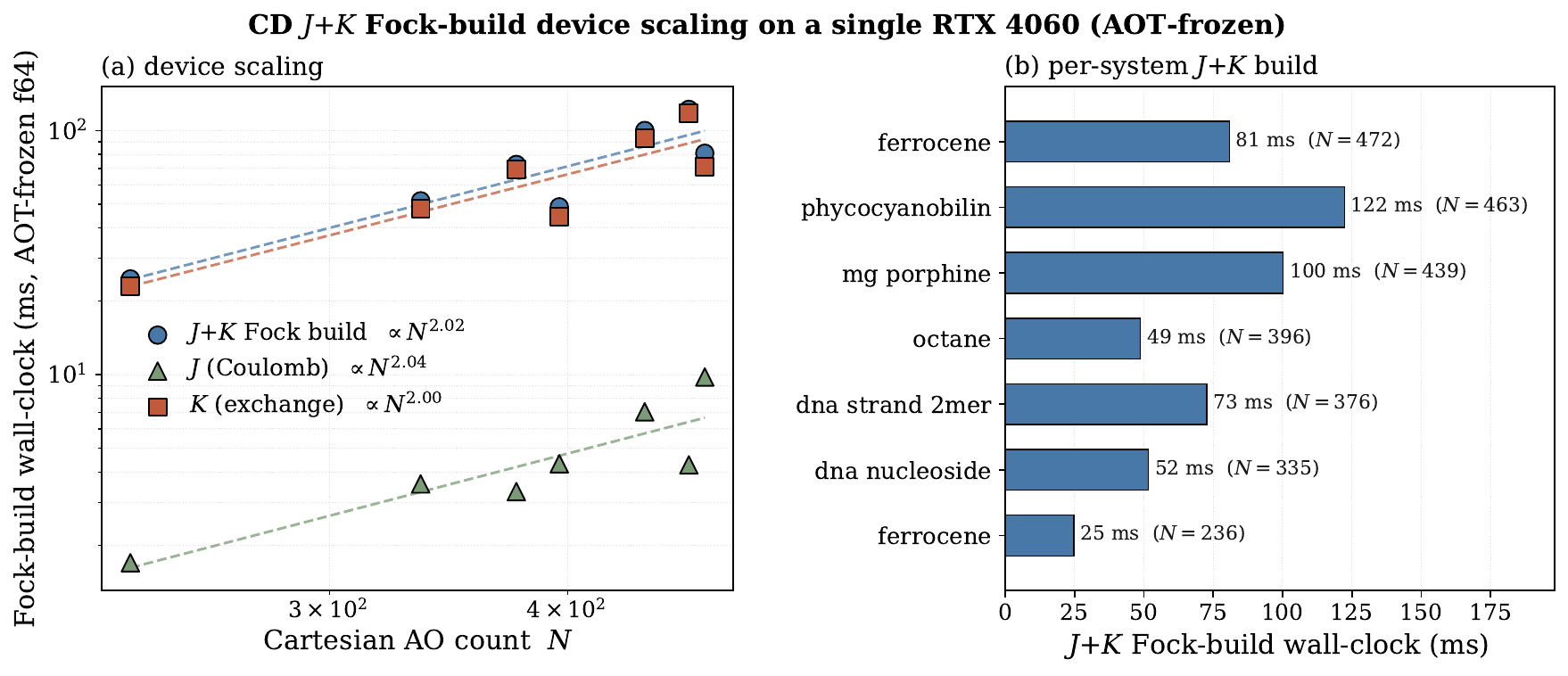}
\caption{Device $J{+}K$ Fock-build performance across the chromophore series on a single
RTX~4060. \textbf{(a)}~Wall-clock versus basis size $N$: $t\propto N^{2.02}$ overall
($J$: $N^{2.04}$; $K$: $N^{2.00}$, both near-quadratic), near the $\sim\!N^{2.0}$
of the efficient hh-TDA of Ref.~\cite{Bannwarth2020}. \textbf{(b)}~Per-system $J{+}K$
Fock-build wall-clock, annotated with each system's Cartesian AO count $N$. The series spans
three basis sets of differing maximum angular momentum, so the exponent is a practical
size-scaling, not a fixed-basis asymptote.}
\label{fig:scaling-series}
\end{figure*}

A profile-guided optimization then attacks the derivative-Fock build itself. Nsight
Compute attribution showed the dominant cost was the two-center metric term of the
exchange channel, built as $6.5\times10^{4}$ single-column kernel launches each at
$\sim\!2\%$ streaming-multiprocessor throughput: the card appeared busy while
$\sim\!98\%$ of its throughput sat idle. Rebuilding that tensor in one full-occupancy
pass ($196$ launches at $55$--$72\%$ throughput), fusing two kernels that redundantly
evaluated the same vertical recurrence, and deriving one channel from the other by a
matrix product together take the warm build from $83.7$\,s to $21.4$\,s---a further
$3.9\times$. Figure~\ref{fig:roofline} shows the throughput recovery behind that
gain---the streaming-multiprocessor occupancy climbing from $\sim\!2\%$ to
$55$--$72\%$ as $65{,}268$ micro-launches collapse to $196$---and Fig.~\ref{fig:speedup}
the resulting three-stage wall-clock cascade, $\sim\!133\times$ cumulative over the
per-pair reference. Every optimization is bit-identical:
in double precision the full $2J-a_xK$ derivative-Fock matrix agrees with the per-pair
reference to $2.2\times10^{-13}$ at $f$-shell ($L=3$), block-independent to
$5\times10^{-14}$; the production single-precision substrate floor is $\sim\!10^{-5}$,
inside the $\sim\!10^{-4}$ achieved analytic-vs-oracle agreement---itself an order of magnitude
inside the $10^{-3}$ acceptance gate (Table~\ref{tab:precision})---and the
assembled NACME output is byte-identical run to run. Table~\ref{tab:precision} reads as
the safety margin of the optimization: the double-precision substrate certifies that the
launch collapse changed only the kernel's launch geometry and not its arithmetic, while
the single-precision production floor sits an order of magnitude inside the tolerance the
physics actually demands. That single-kernel tuning is only half the performance story;
we now step back from the one optimized kernel to the whole production
Cholesky-decomposed (CD) $J/K$ engine, measured across a set of real chromophores
(Figs.~\ref{fig:speedup-series}--\ref{fig:scaling-series}).

\begin{table}[htbp]\centering\small
\caption{Correctness of the split derivative-Fock kernel versus the per-pair reference,
across optimization stages, at $f$-shell (\ce{H2O}, cc-pVTZ, $L=3$). Double-precision
substrate is machine-exact; block-independence confirms the aux-index blocking is a pure
memory knob. The three coupling tolerances are nested: the single-precision floor
($\sim\!10^{-5}$) sits inside the achieved analytic-vs-oracle agreement ($\sim\!10^{-4}$),
which sits inside the $10^{-3}$ acceptance gate.}
\label{tab:precision}
\begin{tabular}{lcc}
\toprule
substrate / check & rel.\ error & block-indep.\\
\midrule
fp64, full $2J-a_xK$ & $2.2\times10^{-13}$ & $5\times10^{-14}$\\
fp32 factor (production) & $\sim\!10^{-5}$ (floor) & $6\times10^{-14}$\\
achieved oracle agreement & $\sim\!10^{-4}$ & ---\\
acceptance gate (coupling) & $10^{-3}$ & ---\\
\bottomrule
\end{tabular}
\end{table}

Figure~\ref{fig:speedup-series} reports the practical speedup of the device CD $J{+}K$
Fock build over a conventional single-thread PySCF FP64 four-index \texttt{get\_jk}, at
the same xTB-optimized geometry and converged density, for seven calculations across six
systems---five chromophores plus an alkane $f$-shell reference---spanning
$236$--$472$ atomic orbitals and three basis sets (def2-SVP, pcseg-0, def2-TZVP). The
device build is $32$--$849\times$ faster (geometric mean $313\times$); across the larger
systems ($\ge\!335$ AO) the range is $323$--$849\times$. The floor of the range is
ferrocene at def2-SVP ($32\times$, $236$ AO, $f$-shell $L=3$), where the CPU reference is
only $792$\,ms and the small absolute work leaves little to amortize; the advantage grows
with system size---the same molecule at def2-TZVP ($472$ AO) reaches $849\times$. Two
disclosures keep this honest. First, the device CD $J/K$ reproduces PySCF's \emph{exact}
four-index $J/K$ to $\sim\!10^{-6}$ ($J$) and $\sim\!10^{-5}$ ($K$): this is a
time-to-the-same-result speedup, not an accuracy shortcut. Second, it is a combined
method-and-hardware advantage---Cholesky-decomposed ERIs on the GPU against conventional
four-index ERIs on the CPU---and is the speedup a user actually obtains, not a
pure-hardware claim; the baseline is conventional PySCF (not density-fitted),
single-threaded (a verified $1$-vs-$6$-thread control gives only $4.8$--$6.0\times$), and
the device $K$ runs in FP64, so the reported factors are a conservative lower bound.

Figure~\ref{fig:roofline-series} places the twelve production CD-DFT kernels on the
RTX~4060 (AD107, sm\_89) roofline---the eight two-electron kernels and the four XC-grid
kernels of Fig.~\ref{fig:newkernels}(b)---showing that the engine is well-placed against the
hardware ceilings across its \emph{whole} footprint, not only at the single kernel of
Fig.~\ref{fig:roofline}. The two-electron kernels populate the bandwidth region: the exchange
($K$) kernels ride $67$--$80\%$ of the FP32 DRAM-bandwidth roof; the Coulomb ($J$) build is
bandwidth-bound at $\sim\!94\%$; and the FP64 $\Gamma$-density build reaches $82\%$ of the
FP64 compute peak---the compute wall that motivates the mixed-precision (FP32) production
substrate. The four grid-based XC kernels sit instead in the compute region at high
arithmetic intensity, as expected for grid-pointwise work: the FP32 $V_{xc}$ scatter and
$f_{xc}$ response reach $6\%$ and $14\%$ of the FP32 roof, and the mostly-FP64 $g_{xc}$
transition-density force and SSF weight-response reach $37\%$ and $16\%$ of the FP64 peak
(the $g_{xc}$ force being the strongest FP64 fraction, saturating the scalar FP64 pipe).
Each kernel sits against its relevant ceiling, so the measured throughput is close to what
the hardware allows rather than an artifact of an under-resolved workload.

Figure~\ref{fig:scaling-series} fits the device $J{+}K$ wall-clock across the series to
$t\propto N^{2.02}$ ($J$: $N^{2.04}$; $K$: $N^{2.00}$, both near-quadratic),
close to the $\sim\!N^{2.0}$ reported for the efficient hh-TDA of Bannwarth and
co-workers~\cite{Bannwarth2020}. Because the series deliberately spans three basis sets of
differing maximum angular momentum, this exponent is a practical size-scaling of the
production workload, not a fixed-basis asymptote. All series numbers were collected under
the ahead-of-time-compiled kernel library with zero just-in-time compilation (generator
modules $0$, new cubins $0$), and the $f$-shell case (ferrocene, $L=3$) is included per
our standing gate rule, so the $f$-shell blind spot is exercised at production scale.

These performance numbers are for the shared two-electron engine---the CD $J/K$ Fock
build through which every method in this work is expressed. The excited-state objects
that headline the paper---the \hhTDA/\ppTDA gradients and NACMEs and the double-hybrid
derivative coupling---now run \emph{device-resident} and AO-direct on this same engine
(Table~\ref{tab:firstofkind}): the frame-covariant numerators, the relaxed-density gradient
assembly, and the double-hybrid response are assembled on the device without forming a
four-index host tensor, each validated against its determinant- or finite-difference oracle.
We nonetheless quote \emph{no} end-to-end device wall-clock for these paths: the
range-separated ($K_{\mathrm{LR}}$) kernels they invoke are not yet captured in the
ahead-of-time-compiled library, so a codegen-free \texttt{aot\_frozen} end-to-end time---the
only kind our standing rule permits---awaits that infrastructure extension. The device
\emph{cost} meanwhile reduces to the \texttt{aot\_frozen} CD $J/K$/$2e$-gradient primitives
timed in the roofline (Fig.~\ref{fig:roofline-series}).

\begin{table*}[tbp]\centering
\caption{Response objects reported here and their device-residency status. All three run
\emph{device-resident} and AO-direct through the shared CD $J/K$ engine---no four-index host
tensor---and each is validated against an independent determinant/finite-difference oracle;
of these, only the double-hybrid CIS(D) coupling is first of its kind, the hh/pp gradient and
NACME being consumer-hardware reproductions of the hybrid-only realizations of
Refs.~\cite{Bannwarth2020,Yu2020}. Each reduces to the CD $J/K$/$2e$-gradient primitives that
are ahead-of-time-compiled and timed in the roofline (Fig.~\ref{fig:roofline-series}). Per our
standing rule, \emph{no end-to-end device wall-clock is quoted}: the range-separated kernels
these paths invoke are not yet in the \texttt{aot\_frozen} library, so a codegen-free
end-to-end time awaits that extension (an honest partial).}
\label{tab:firstofkind}
\begin{tabular}{@{}lll@{}}
\toprule
object & validation & device-residency status (this branch)\\
\midrule
\parbox[t]{0.27\textwidth}{\raggedright\hhTDA/\ppTDA\ excited-state gradient~\cite{Bannwarth2020,Yu2020}\strut} &
\parbox[t]{0.28\textwidth}{\raggedright finite differences, $\le\!10^{-6}\,E_h/a_0$ (Table~\ref{tab:hhpp})\strut} &
\parbox[t]{0.35\textwidth}{\raggedright fully device-resident: eigenvalue gradient \emph{and} relaxed-density assembly, FD-validated\strut}\\[4pt]
\parbox[t]{0.27\textwidth}{\raggedright\hhTDA/\ppTDA\ NACME\strut} &
\parbox[t]{0.28\textwidth}{\raggedright two-hole/two-particle overlap oracle, $\le\!10^{-4}$ (Table~\ref{tab:hhpp})\strut} &
\parbox[t]{0.35\textwidth}{\raggedright device-resident end-to-end---AO-direct frame-covariant numerator $+$ term(c), no host four-index tensor---validated to the overlap oracle including the $f$-shell ($L=3$)\strut}\\[4pt]
\parbox[t]{0.27\textwidth}{\raggedright double-hybrid CIS(D) coupling $d^{(D)}$\strut} &
\parbox[t]{0.28\textwidth}{\raggedright literal determinant-overlap oracle, $\sim\!10^{-4}$ (Table~\ref{tab:oracle})\strut} &
\parbox[t]{0.35\textwidth}{\raggedright device-resident and AO-direct: ERI/C 4-index derivative, CPHF/CIS response $+$ B88/LYP $f_{xc}$, and \texttt{make\_h1} RHS; frozen-reference amplitudes remain host by design\strut}\\
\bottomrule
\end{tabular}
\end{table*}

\section{Conclusions}
\label{sec:conclusions}
A single reverse-mode contraction-DAG engine, closed under one non-symmetric
AO-Laplace $J/K$ primitive and never forming a four-index MO tensor, delivers analytic
nuclear gradients and interstate nonadiabatic couplings across a family of
density-functional excited-state methods---the topology-correct \hhTDA/\ppTDA manifolds
and the non-variational doubles sector of a double-hybrid coupling---validated against
an independent literal-wavefunction oracle, with the gradients, NACMEs, and the double-hybrid
coupling all executed device-resident and AO-direct---through one shared two-electron
engine---within the 8\,GB of a consumer GPU. The construction is uniform: each method's gradient and NACME
is a corollary of the same transpose plus the same kernel, and the hardest object (the
double-hybrid coupling) is reached by the same machinery that gives the topology-correct
manifolds their forces. A profile-guided $\sim\!133\times$ launch collapse, with
double-precision bit-identity preserved, brings the coupling's dominant two-electron kernel
into interactive wall-clock on hardware a student owns, and across a chromophore series of up to
$472$ AOs spanning three basis sets the production CD $J/K$ engine runs $32$--$849\times$ faster than a
conventional four-index CPU reference at the same accuracy. A certified codegen-free
(\texttt{aot\_frozen}) end-to-end wall-clock for the range-separated NACME path---which
requires the ahead-of-time-compiled library to capture the long-range kernels---and coupling
the engine to nonadiabatic dynamics through the correct conical-intersection topology are the
natural next steps.

\section*{Supplementary Material}
The supplementary material provides the full theoretical and validation record behind the
results reported here: the reverse-mode ``relaxation is DAG transpose'' principle and its
shared adjoint facility (Sec.~S2); the AO-direct Laplace representation and the
non-symmetric $J/K$ geometry kernel that interstate responses require (Sec.~S3); the
response templates for the analytic gradients (Sec.~S4); the complete from-first-principles
working equations for the double-hybrid CIS(D) derivative coupling, derived as its three
physical contributions (Sec.~S5); the literal determinant-overlap ($p^{\dagger}q$)
oracles---the four-block CIS(D) wavefunction overlap, the generalized biorthogonal
replacement-determinant formula, the parameter-free $u\!\to\!0$ consistency anchors, and the
per-contribution sub-oracles (Sec.~S6); the numerical validation results (Sec.~S7); the
twisted-ethylene $S_0/S_1$ sign-fixed coupling booking and its determinant-oracle
convergence (Sec.~S8); the electron-translation-factor construction and the per-atom
partition freedom (Sec.~S9); and a reproducibility map of the host-side verification modules
(Sec.~S10). The determinant oracles and the host-side scripts that certify every working
equation on CPU hardware, together with these derivations, are additionally deposited as
described in the Data Availability statement.

\begin{acknowledgments}
We acknowledge financial support and computational resources provided by NeuroTechNet S.A.S.
\end{acknowledgments}

\section*{Author Declarations}
\subsection*{Conflict of Interest}
The author has no conflicts to disclose.

\section*{Data Availability}
The data that support the findings of this study---the literal determinant-overlap
($p^{\dagger}q$) oracles and the host-side scripts that verify every working equation of
the analytic couplings and gradients against them, together with the detailed derivations---
are openly available. They are archived on Zenodo under a permanent DOI and mirrored in a
public GitHub repository under the MIT license, pinned to a tagged commit; the bundle is
provided to reviewers through an anonymized link at submission and released publicly upon
acceptance. The bundle is self-contained: it reproduces every entry of the verification
tables (Tables~\ref{tab:oracle} and~\ref{tab:hhpp}) on CPU hardware from \textsc{PySCF} and
\textsc{NumPy} alone, without access to any GPU code, so that every accuracy claim---which
carries the scientific content of this work---is independently checkable.

\emph{Withheld:} the performance-tuned CUDA $J/K$ and derivative kernels, and the code
generators that emit them (a tuned application of the framework of
Ref.~\cite{GuerreroRECURSUM}), are proprietary to the funder (NeuroTechNet~S.A.S.) and are
available from the corresponding author on reasonable request. The per-system timing,
roofline, and scaling data these kernels produced are released with the bundle, so the
reported speedups can be recomputed from the deposited files, although \emph{re-measuring}
them requires the withheld kernels. We regard this as a real limitation on the
reproducibility of the performance claims only; the correctness claims are unaffected.

\bibliographystyle{aipnum4-2}
\bibliography{refs}

\end{document}